\documentclass[a4paper,fleqn,usenatbib]{mnras}

\usepackage{mathptmx}

\usepackage[T1]{fontenc}
\usepackage{ae,aecompl}

\usepackage{graphicx}	
\usepackage{amsmath}	
\usepackage{amssymb}	
\usepackage[utf8]{inputenc} 
\usepackage{array} 
\usepackage{rotfloat} 
\usepackage{multicol,rotating} 
\usepackage{subcaption} 
\usepackage[section]{placeins} 

\title[Decimetre-scaled observations of a cometary surface]{Decimetre-scaled spectrophotometric properties of the nucleus of comet 67P/Churyumov-Gerasimenko from OSIRIS observations}

\author[Feller et al.]{
\parbox{\textwidth}{
\begin{normalsize}
C. Feller$^{1}$\thanks{E-mail: \texttt{clement.feller<at>obspm.fr} },
S. Fornasier$^{1}$, P.H. Hasselmann$^{1}$, A. Barucci$^{1}$, F. Preusker$^{2}$, F. Scholten$^{2}$, L. Jorda$^{3}$, 
A. Pommerol$^{4}$, B. Jost$^{4}$, O. Poch$^{4}$, M.R. ElMaary$^{4}$, N. Thomas$^{4}$, I. Belskaya$^{5}$, M. Pajola$^{6,7}$,
H. Sierks$^{8}$, C. Barbieri$^{9}$, P.L. Lamy$^{10}$, D. Koschny$^{11}$, H. Rickman$^{12,13}$, R. Rodrigo$^{14,15}$,
J. Agarwal$^{8}$,
M. A'Hearn$^{16}$,
J.-L. Bertaux$^{17}$,
I. Bertini$^{7}$,
S. Boudreault$^{8}$,
G. Cremonese$^{18}$,
V. Da Deppo$^{19}$,
B.J.R. Davidsson$^{20}$,
S. Debei$^{21}$,
M. De Cecco$^{22}$,
J. Deller$^{8}$,
M. Fulle$^{23}$,
A. Giquel$^{8}$,
O. Groussin$^{3}$,
P.J. Gutierrez$^{24}$,
C. G\"uttler$^{8}$,
M. Hofmann$^{8}$,
S.F. Hviid$^{2}$,
H. Keller$^{25}$,
W.-H. Ip$^{26}$,
J. Knollenberg$^{2}$,
G. Kovacs$^{8}$,
J.-R. Kramm$^{8}$,
E. K\"uhrt$^{2}$,
M. K\"uppers$^{27}$,
M. L. Lara$^{25}$,
M. Lazzarin$^{9}$,
C. Leyrat$^{1}$,
J.J. Lopez Moreno$^{25}$,
F. Marzari$^{9}$,
N. Masoumzadeh$^{8}$,
S. Mottola$^{2}$,
G. Naletto$^{6,20,28}$,
N. Oklay$^{8}$,
X. Shi$^{8}$,
C. Tubiana$^{8}$,
J.-B. Vincent$^{8}$
\end{normalsize}} 
\\~\\
\parbox{\textwidth}{
$^{1}$LESIA, Observatoire de Paris, PSL Research University, CNRS, Univ. Paris Diderot, Sorbonne Paris Cité, UPMC Univ., Paris 06, Sorbonne Université, 5 Place J. Janssen, Meudon Cedex 92195, France;
$^{2}$Deutsches Zentrum fuer Luft- und Raumfahrt (DLR), Institut fuer Planetenforschung, Asteroiden und Kometen, Rutherfordstrasse 2, 12489 Berlin, Germany;
$^{3}$Aix-Marseille Université, CNRS, LAM (Laboratoire d'Astrophysique de Marseille), UMR 7326, 38 rue Frédéric Joliot-Curie, 13388 Marseille, France;
$^{4}$Physikalisches Institut der Universitaet Bern, Sidlerstr. 5, 3012 Bern, Switzerland;
$^{5}$Institute of Astronomy,V.N. Karazin National University, Sumska Str. 35, Kharkiv 61022, Ukraine;
$^{6}$NASA Ames Research Center, Moett Field, CA 94035, USA; 
$^{7}$Center of Studies and Activities for Space (CISAS) “G. Colombo”, University of Padova, Via Venezia 15, 35131 Padova, Italy
$^{8}$Max-Planck-Institut für Sonnensystemforschung, Justus-von-Liebig-Weg, 3,  37077, Goettingen, Germany;
$^{9}$University of Padova, Department of Physics and Astronomy “Galileo Galilei”, Vicolo dell'Osservatorio 3, 35122 Padova, Italy;
$^{10}$Aix-Marseille Université, CNRS, LAM (Laboratoire d'Astrophysique de Marseille), UMR 7326, 38 rue Frédéric Joliot-Curie, 13388 Marseille, France;
$^{11}$Research and Scientific Support Office, European Space Research and Technology Centre/ESA, Keplerlaan 1, Postbus 299, 2201 AZ Noordwijk ZH, The Netherlands;  
$^{12}$PAS Space Research Center, Bartycka 18A, 00716, Warszawa, Poland;  
$^{13}$Dept. of Physics and Astronomy, Uppsala University, Uppsala, Sweden;
$^{14}$Centro de Astrobiologia (INTA -CSIC), European Space Agency, European Space Astronomy Centre (ESAC), P.O. box 78, E-28691, Villanueva de la Canada, Madrid, Spain;   
$^{15}$International Space Science Institute, Hallerstrasse 6, 3012 Bern, Switzerland;  
$^{16}$University of Maryland, Department of Astronomy, College Park, MD 20742-2421, USA;  
$^{17}$LATMOS, CNRS/UVSQ/IPSL, 11 boulevard d'Alembert,78280, Guyancourt, France;  
$^{18}$INAF, Osservatorio Astronomico di Padova, Vicolo dell'Osservatorio 5, 35122 Padova, Italy; 
$^{19}$CNR-IFN UOS Padova LUXOR, Via Trasea, 7, 35131 Padova, Italy; 
$^{20}$Jet Propulsion Laboratory, M/S 183-301, 4800 Oak Grove Drive, Pasadena, CA 91109, U.S.A.
$^{21}$Department of Industrial Engineering - University of Padova, via Venezia 1, 35131 Padova, Italy; 
$^{22}$University of Trento, Via Mesiano 77, 38100 Trento, Italy; 
$^{23}$INAF - Osservatorio Astronomico, Via Tiepolo 11, 34014 Trieste, Italy; 
$^{24}$Instituto de Astrofisica de Andalucia (CSIC), c/ Glorieta de la Astronom\'ia s/n, 18008 Granada, Spain; 
$^{25}$Institut fuer Geophysik und extraterrestrische Physik (IGEP), Technische Universität Braunschweig, Mendelssohnstr. 3, 38106 Braunschweig, Germany; 
$^{26}$National Central University, Graduate Institute of Astronomy, 300 Chung-Da Rd, Chung-Li 32054 Taiwan; 
$^{27}$ESA/ESAC, P.O. box 78, E-28691, Villanueva de la Canada, Madrid, Spain; 
$^{28}$University of Padova, Department of Information Engineering, Via Gradenigo 6/B, 35131 Padova, Italy}
} 

\date{Accepted XXX. Received YYY; in original form ZZZ}

\pubyear{2016}

\begin{document}
\label{firstpage}
\pagerange{\pageref{firstpage}--\pageref{lastpage}}
	\maketitle

\begin{abstract}
We present the results of the photometric and spectrophotometric properties of the 
67P/Churyumov-Gerasimenko nucleus derived with the OSIRIS instrument during the 
closest fly-by over the comet, which took place on 14$^{th}$ February 2015 at a 
distance of $\sim$ 6 km from the surface. Several images covering the 0--33$^{\circ}$
phase angle range were acquired, and the spatial resolution achieved was 11 cm/pxl. 
The flown-by region is located on the big lobe of the comet, near the borders of 
the Ash, Apis and Imhotep regions.
Our analysis shows that this region features local heterogeneities at the decimetre 
scale. We observed difference of reflectance up to 40\%  between bright spots 
and sombre regions, and spectral slope variations up to 50\%. The spectral 
reddening effect observed globally on the comet surface by 
\citet{Fornasier_2015_AA_67PCG} is also observed locally on this region, but 
with a less steep behaviour. We note that numerous metre-sized boulders, which 
exhibit a smaller opposition effect, also appear spectrally redder than their 
surroundings.  In this region, we found no evidence linking observed bright spots 
to exposed water-ice-rich material. We fitted our dataset using the Hapke 2008 
photometric model. The region overflown is globally as dark as the whole nucleus 
(geometric albedo of 6.8\%) and it has a high porosity value in the uppermost-layers (86\%). 
These results of the photometric analysis at a decimetre scale indicate that the 
photometric properties of the flown-by region are similar to those previously 
found for the whole nucleus.
\end{abstract}



\section{Introduction}
The ROSETTA mission is the cornerstone mission of the European Space Agency 
devoted to the study of the minor bodies of the Solar System. Its primary 
objective is to perform an extensive study of the comet 67P/Churyumov-Gerasimenko 
(hereafter 67P/CG), of its nucleus as well as the contiguous cometary environment 
with the help of several instruments, including cameras, spectrometers, dust 
analysers and radio science experiments.

Previous space missions designed for comets exploration have helped building up an awesome 
trove of knowledge of those objects, yet the spacecrafts only performed flyby manoeuvres 
(we refer the reader to \cite{AHearn_2005_Science_DeepImpactResults}, \cite{AHearn_2011_Science_EpoxiMission},  \cite{2011A&ARv..19...48B}, 
and references therein). The Rosetta mission is the first one to escort a comet from 
4 AU inbound, through its perihelion (1.24 AU in the case of 67P/CG) and back 
to 3 AU outbound, as well as the first one to have executed a flyby manoeuvre 
about 6 km above a cometary nucleus. 

The Optical, Spectroscopic, and Infrared Remote Imaging System (OSIRIS) instrument 
is the scientific camera system on-board the ROSETTA orbiter. 

This imaging system has allowed to perform extensive studies of the 67P/CG nucleus. 
Beyond depicting a peculiar bi-lobate shape, OSIRIS has characterized a surface 
exhibiting a complex morphology, including both fragile and consolidated rocky terrains, 
dusty regions, depressions and pits, sometime active, and extensive layering 
\citep{Sierks_2015_Science_67PCG, Thomas_2015_AA_Dust_ReDeposition, Vincent_2015_Nature_ActivePits, Massironi_2015_Nature_67PCG_bilobate}. 
Photometric and spectrophotometric properties of the 67P/CG comet were studied by 
\cite{Fornasier_2015_AA_67PCG,LaForgia_2015_AA_Agilkia,Oklay_2016_AA_Variegation, Lucchetti_2016_AA_Abydos, Pajola_2016_AA_67P_Aswan_review}. 
In \cite{Fornasier_2015_AA_67PCG}, the authors reported a globally red spectral 
behaviour for the nucleus, consistent with that of primitive Solar System bodies 
like Jupiter Trojans \citep{Fornasier_2004_Icarus_Trojans, Fornasier_2007_Icarus_TrojansSurvey} 
and part of the trans-neptunian population \citep{Fornasier_2009_AA_TNOCentaurSurvey}.
They found colour and albedo heterogeneities over the nucleus. Three kind of 
terrains were identified on the basis of the spectral slope, those having a bluer 
(i.e. a less steep spectral slope value) being associated to areas enriched in water-ice 
content, and those having a higher spectral slope value were found predominantly 
to correspond to dust rich regions. \\

\begin{figure}
\centering
   \includegraphics[width=0.49\textwidth]{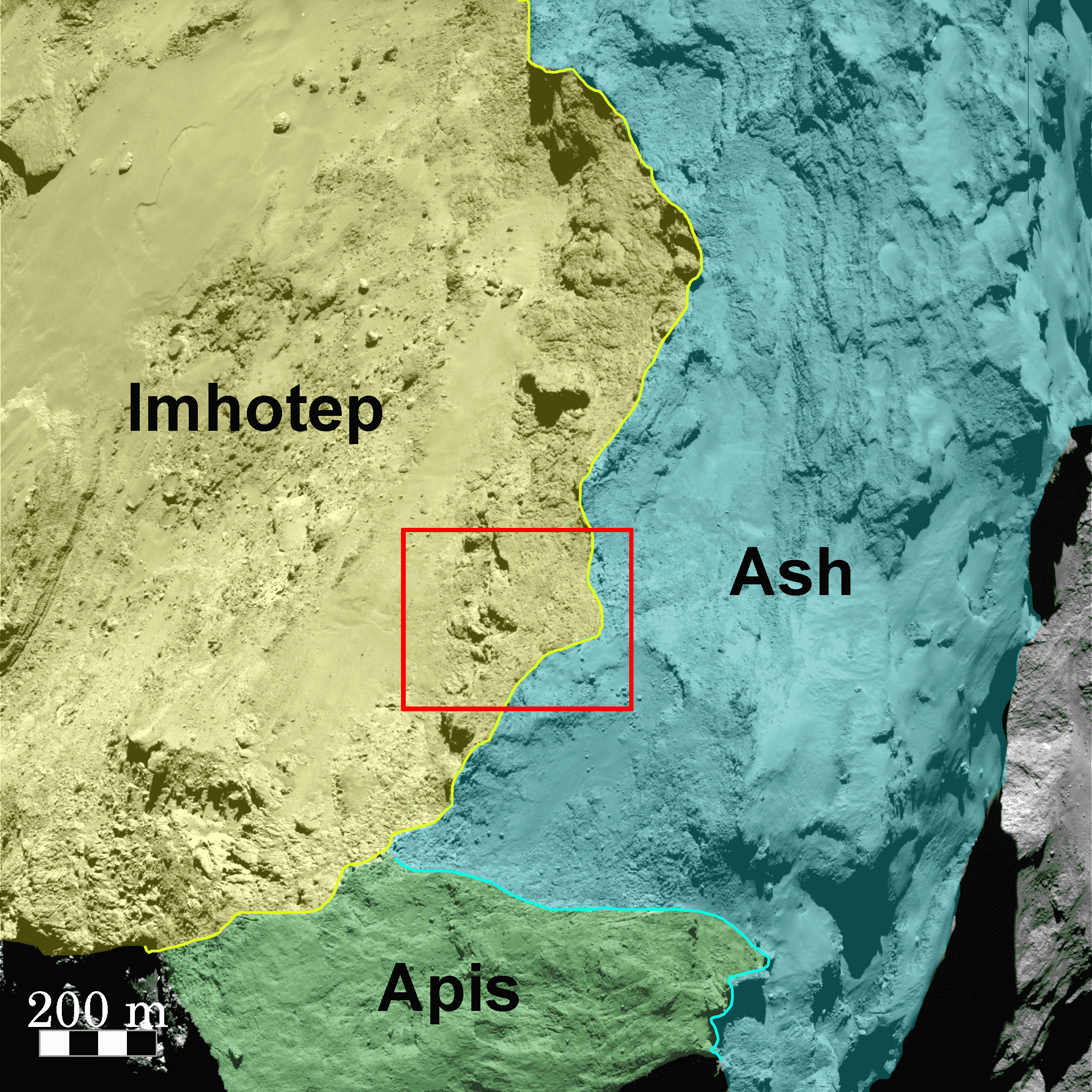}
   \caption{WAC image taken on the 14$^{th}$ of February 2015 at UTC - 14:36:55 depicting the flyby area (red square) and the surroundings morphological regions. The scale in the middle of the image is of 0.92 metre per pixel. The dimensions of the red square are about 350 metres per 225 metres.}
    \label{Fig:ContextFrame}
\end{figure}

The comet is dark: its geometric albedo reaches 6.5 \% at 649 nm, and at a decametre scale,
we observed local reflectance variations up to ~25\% \citep{Fornasier_2015_AA_67PCG}.  The VIRTIS 
spectrometer results indicate a global surface composition dominated by 
dehydrated and organic-rich refractory materials \citep{Capaccioni_2015_Science_Virtis}. 
However, several local bright spots were identified in the OSIRIS images 
\citep{Pommerol_2015_AA_ExposedIce, Fornasier_2016_Science_IceDust, Barucci_2016_AA_Ices_Features, Oklay_2016_AA_Variegation} 
and interpreted as exposures of water-ice. The presence of water-ice has been 
confirmed by the VIRTIS observations in the Hapi and Imhothep regions, and in 
some of the bright spots \cite{DeSanctis_2015_Nature_WaterCycle, Filacchione_2016_Icarus_GlobalSurfaceVirtis, Barucci_2016_AA_Ices_Features}.

In this work, we investigate the photometric and spectro-photometric properties 
of the comet 67P/CG's nucleus based on a dataset of images, taken by the OSIRIS 
imaging system, with the best spatial resolution (11 cm/pxl) since the S/C had started 
to escort the comet, and in which the phase angle ranges from 0$^{\circ}$ up to 
33$^{\circ}$.
We sought thus to study the heterogeneity of the nucleus at the decimetre 
scale and to constraint parameters modelling the surge of intensity of the 
reflected light at low phase angles.\\
The following section presents the instrument and the observations in question, and the third one 
summarizes the steps of data reduction and preparation.
In the fourth and fifth sections, we shall present the results of our spectrophotometric analyses examining, 
first, the flown-by region as a whole, and then considering parts of this region. 
After which we shall present the results of our photometric analysis in the sixth section.
We discuss the results of our analyses in the seventh section.

 \section{Observations}
 
\begin{figure}
     \includegraphics[scale=0.25]{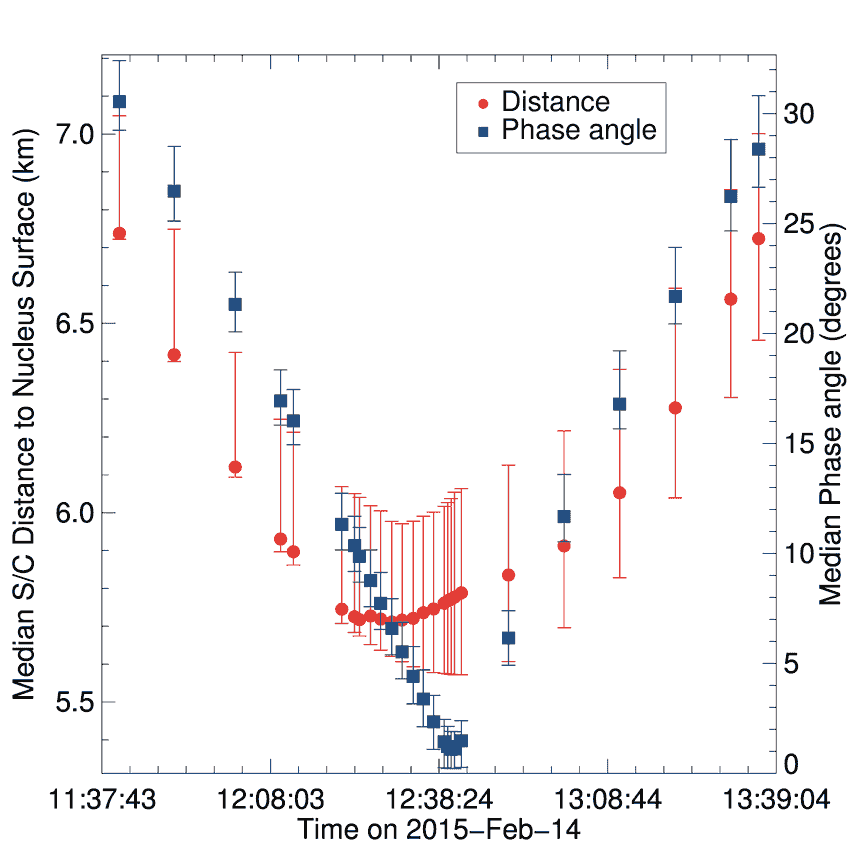}
          \caption{Variations of the median phase angle and distance of the spacecraft to the surface variations during the flyby. The bars around the median values correspond to the computed extrema according to the 3D model.}
          \label{fig:DistPhaTime}
\end{figure}

\begin{figure}
	\includegraphics[width=0.49\textwidth]{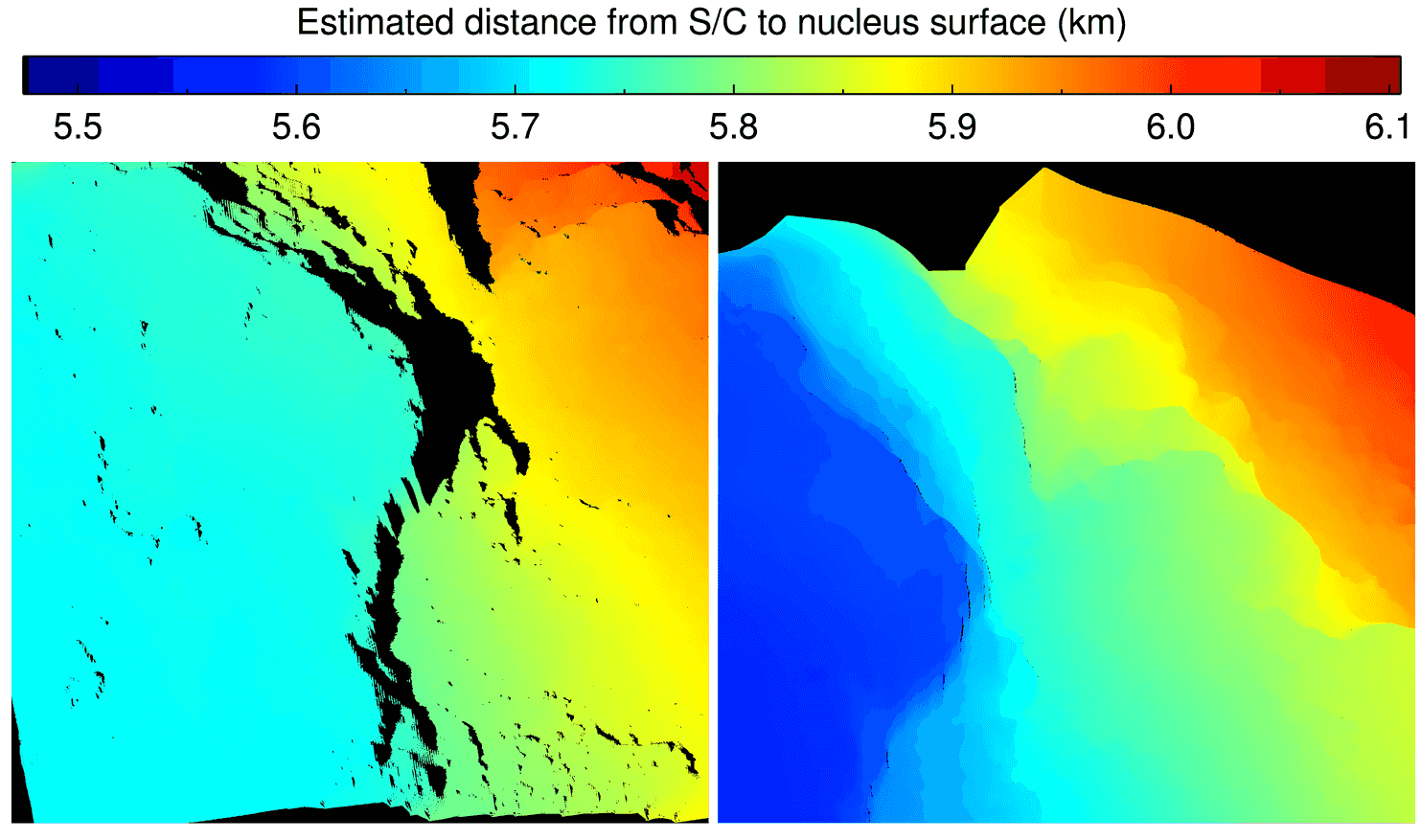}
	\caption{Computed depth maps with the 3D model at UTC - 12:20:54 (left) and at UTC - 12:39:25 (right) as perceived from the Rosetta spacecraft. The black areas correspond either to shadows or to regions not defined in the 3D model of the region.}
    \label{fig:DepthMaps}
\end{figure}

\begin{table}
  \centering
  \begin{tabular}{|c|c|c|}
	\hline
	NAC Filter & Wavelength & Bandwidth \\
        &  [nm]    & [nm] \\
	\hline
	\hline
 F15 & 269.3 & 53.6\\
 F16 & 360.0 & 51.1\\
 F84 & 480.7 & 74.9\\
 F83 & 535.7 & 62.4\\
 F82 & 649.2 & 84.5\\
 F27 & 701.2 & 22.1\\
 F88 & 743.7 & 64.1\\
 F51 & 805.3 & 40.5\\
 F41 & 882.1 & 65.9\\
 F61 & 931.9 & 34.9\\
 F71 & 989.3 & 38.2\\
	\hline
  \end{tabular}
  \caption{Table listing the OSIRIS filters' name, with the associated central wavelength and bandwidth. }
  \label{Filters List}
\end{table}

The OSIRIS imaging system comprises two cameras: the Narrow-Angle Camera 
(hereafter NAC) and the Wide-Angle Camera (hereafter WAC). The NAC has a 
field-of-view of 2.35$^{\circ}$x2.35$^{\circ}$, it was designed to perform 
observations of the nucleus with broad-band filters ranging from the near UV to 
the near IR domains (240-1000 nm) at an angular resolution of 18.6 $\mu$rad per 
pixel. Those filters are optimized for the study of the nucleus' mineralogy. 
The WAC has a field-of-view of 11.6$^{\circ}$x12.1$^{\circ}$, and was designed 
to perform observations of the dust and gaseous species in the coma in the same 
wavelength domain (250-1000 nm) though with narrow-band filters at an 
angular resolution of 101 $\mu$rad per pixel.\\
Both cameras are equipped with the same detector system: a 2048x2048 pixels 
array CCD with a pixel size of 13.5 $\mu$m. For more detailed description of 
the instrument, we refer the reader to \cite{Keller_2007_SSR_Osiris}.\\

 On the 14$^{th}$ of February 2015, the ROSETTA spacecraft flew over the 
Imhotep region, and the closest approach took place close to the boundary between Ash, 
Apis and Imhotep, as depicted in Fig. \ref{Fig:ContextFrame}.
For detailed descriptions of the definition of the comet's morphological and 
structural regions, we refer the reader to \cite{Thomas_2015_Science_67P_morphology}
and \cite{El-Maarry_2015_AA_67PCG_GlobalMorphology}.\\

Around the time of closest approach, the OSIRIS instrument took series of 
observations using several filters with both NAC and WAC cameras. For the NAC, 
those observations were performed using two sets of filters: 23 series of 
observations with 3 filters (namely F82, F84 and F88), and 2 series of 
observations with all the 11 NAC filters (see Tables \ref{Filters List}, and 
\ref{Image List}).\\
The WAC took series of observations using only 2 filters (one centred at 375.6 
nm, another at 631.6 nm). However, as multiple shutter errors plagued the 
WAC during the flyby, we did not include those images in our 
analysis.\\

At that epoch, the comet was 2.31 AU away from the Sun and according to the 
trajectory and attitude reconstruction in the frame of the Cheops boulder 
\citep{Preusker_2015_AA_67P_SPG}, the median distance between the spacecraft 
and the part of the nucleus surface apparent in the NAC field-of-view varies between 
5.711$^{+0.265}_{-0.090}$ km and 6.737$^{+0.310}_{-0.015}$ km.
The minimum distance was attained at 12:39:54 (Fig. \ref{fig:DistPhaTime}). 
The variations of the uncertainties displayed in this figure are due to the 
dispersion of the values as the camera's field-of-view pans different parts 
of the flown-by region. The median phase angle in the frame of the CCD varies 
between $1.09^{\circ}~^{+0.82^{\circ}}_{-1.09^{\circ}}$ and 
30.5$^{\circ} ~ ^{+1.9^{\circ}}_{-1.3^{\circ}}$.
The NAC field of view being 2.35$^{\circ}$x2.35$^{\circ}$ wide, the comet surface was observed 
with a decimetre resolution and at a phase angle of zero between 12:39:58 (UTC) 
and 12:42:25 (UTC), as confirmed by the presence of the penumbra of Rosetta in 
the corresponding images (see Fig. \ref{fig:Image NAC F22 ContextFrame 2}).

The region flown-by has been identified as a part of the boundary between the 
morphological regions Ash and Imhotep. A dedicated 3D model of the region has 
been generated from this dataset \citep{Preusker_2015_EPSC_67P_Model}. In the 
Cheops frame, this region extends from $-170.12^{\circ}$ to $178.99^{\circ}$ in longitude,
from $5.41^{\circ}$ to $14.31^{\circ}$ in latitude, and from $1.954$ km to $2.445$ km 
in distance from the nucleus centre of mass.

This analysed area is a transition region located between the fine deposits on 
the layered terrain unit (distance $\leq$5.75 km, in Fig. \ref{fig:DepthMaps}, left image) 
and the outcropping layered terrain unit (distance $\geq$ 5.7 km, in Fig. \ref{fig:DepthMaps}, right image) 
presented in \cite{Giacomini_2016_Icarus}. The clear and unambiguous 
morphological limit which is here present, is the main reason for the distinction 
between the two separate geographical regions, i.e. the Ash region and the Imhotep 
depression \citep{El-Maarry_2015_AA_67PCG_GlobalMorphology}. As from \citet{Massironi_2015_Nature_67PCG_bilobate} 
(see the extended data Fig. 2e) the studied area clearly shows the presence of 
metre to centimetre scale strata heads pointing towards Imhotep. Such strata are 
almost continuously present in the 500m steep cliff, demonstrating that the 
proposed layered structure of 67P/CG is not only detectable on a global scale, but 
also on the highest resolution images.

\section{Data reduction}

\begin{table*}
\centering
\begin{tabular}{|c|c|c|c|c|c|}
\hline
Camera & Time & Filter & Phase Angle & S/C Distance to & Resolution\\
              &(UTC)&            & ($^{\circ}$)   & Nucleus Surface (km)  & (m/pxl)  \\
\hline
\hline
NAC & 2015-02-14T11:40:54 & F82, F84, F88  & 30.5 $\pm$ 1.4 & 6.73 $\pm$ 0.05 & 0.126 $\pm$ 0.001\\
NAC & 2015-02-14T11:50:45 & F82, F84, F88  & 26.4 $\pm$ 1.4 & 6.41 $\pm$ 0.07 & 0.120 $\pm$ 0.001\\
NAC & 2015-02-14T12:01:44 & F82, F84, F88  & 21.3 $\pm$ 1.2 & 6.12 $\pm$ 0.08 & 0.115 $\pm$ 0.001\\
NAC & 2015-02-14T12:09:54 & F15, F16, F82, F83, & 16.9 $\pm$ 1.1 & 5.93 $\pm$ 0.08 & 0.111 $\pm$ 0.001\\
 - & - & F84, F27, F88, F41, &  - &  - &  - \\
 - & - & F61, F71, F51  &  - &  - &  - \\
NAC & 2015-02-14T12:12:12 & F82, F84, F88  & 16.0 $\pm$ 1.1 & 5.89 $\pm$ 0.08 & 0.111 $\pm$ 0.001\\
NAC & 2015-02-14T12:20:54 & F15, F16, F82, F83, & 11.3 $\pm$ 1.1 & 5.74 $\pm$ 0.08 & 0.108 $\pm$ 
0.001\\
       - &										- & F84, F27, F88, F41, &  - &  - &  - \\
       - & 										- & F61, F71, F51  &	- &  - &  - \\
NAC & 2015-02-14T12:23:12 & F82, F84, F88  & 10.3 $\pm$ 1.0 & 5.72 $\pm$ 0.08 & 0.107 $\pm$ 0.001\\
NAC & 2015-02-14T12:24:05 & F82, F84, F88  & 9.8 $\pm$ 1.0 & 5.71 $\pm$ 0.08   & 0.107 $\pm$ 
0.001\\
NAC & 2015-02-14T12:26:05 & F82, F84, F88  & 8.7 $\pm$ 1.0 & 5.72 $\pm$ 0.08   & 0.107 $\pm$ 
0.001\\
NAC & 2015-02-14T12:27:54 & F82, F84, F88  & 7.7 $\pm$ 1.0 & 5.71 $\pm$ 0.09 & 0.107 $\pm$ 0.001\\
NAC & 2015-02-14T12:29:54 & F82, F84, F88  & 6.6 $\pm$ 0.9 & 5.71 $\pm$ 0.09 & 0.107 $\pm$ 0.001\\
NAC & 2015-02-14T12:31:45 & F82, F84, F88  & 5.5 $\pm$ 0.9 & 5.71 $\pm$ 0.10 & 0.107 $\pm$ 0.002\\
NAC & 2015-02-14T12:33:45 & F82, F84, F88  & 4.4 $\pm$ 0.8 & 5.72 $\pm$ 0.11 & 0.107 $\pm$ 0.002\\
NAC & 2015-02-14T12:35:34 & F82, F84, F88  & 3.3 $\pm$ 0.8 & 5.73 $\pm$ 0.11 & 0.107 $\pm$ 0.002\\
NAC & 2015-02-14T12:37:25 & F82, F84, F88  & 2.3 $\pm$ 0.7 & 5.74 $\pm$ 0.12 & 0.108 $\pm$ 0.002\\
NAC & 2015-02-14T12:39:19 & F82, F84, F88  & 1.4 $\pm$ 0.6 & 5.76 $\pm$ 0.12 & 0.108 $\pm$ 0.002\\
NAC & 2015-02-14T12:39:58 & F82, F84, F88  & 1.2 $\pm$ 0.5 & 5.76 $\pm$ 0.12 & 0.108 $\pm$ 0.002\\
NAC & 2015-02-14T12:40:34 & F82, F84, F88  & 1.0 $\pm$ 0.5 & 5.77 $\pm$ 0.12 & 0.108 $\pm$ 0.002\\
NAC & 2015-02-14T12:41:11 & F82, F84           & 1.1 $\pm$ 0.5 & 5.77 $\pm$ 0.12 & 0.108 $\pm$ 0.002\\
NAC & 2015-02-14T12:42:25 & F82                   & 1.4 $\pm$ 0.6 & 5.78 $\pm$ 0.12 & 0.108 $\pm$ 0.002\\
NAC & 2015-02-14T12:50:54 & F82, F84, F88  & 6.1 $\pm$ 0.8 & 5.83 $\pm$ 0.12 & 0.109 $\pm$ 0.002\\
NAC & 2015-02-14T13:00:54 & F82, F84, F88  & 11.6 $\pm$ 1.3 & 5.91 $\pm$ 0.11 & 0.111 $\pm$ 0.002\\
NAC & 2015-02-14T13:10:54 & F82, F84, F88  & 16.7 $\pm$ 1.6 & 6.05 $\pm$ 0.12 & 0.113 $\pm$ 0.002\\
NAC & 2015-02-14T13:20:54 & F82, F84, F88  & 21.6 $\pm$ 1.6 & 6.27 $\pm$ 0.13 & 0.118 $\pm$ 0.002\\
NAC & 2015-02-14T13:30:54 & F82, F84, F88  & 26.2 $\pm$ 1.8 & 6.56 $\pm$ 0.13 & 0.123 $\pm$ 0.002\\
NAC & 2015-02-14T13:35:54 & F82, F84, F88  & 28.3 $\pm$ 1.9 & 6.72 $\pm$ 0.12 & 0.126 $\pm$ 0.002\\
\hline
\end{tabular}
\caption{List of the NAC observations analysed. The time column corresponds to 
the epoch at which the first observation of the sequence was taken. The values 
indicated in the last 3 columns correspond to the median values and their 
dispersion over the NAC field-of-view computed while producing simulations of 
the NAC observations with a specific 3D shape model of the region flown-by 
produced by the DLR.}
\label{Image List}
\end{table*}

The images used for this study were reduced using the OSIRIS standard pipeline 
(OsiCalliope v.1.0.0.21) up to level 3B, following the data reduction steps 
described in \cite{Kueppers_2007_AA_Steins_lightcurve}, and \cite{Tubiana_2015_AA_OsirisQuality}.\\
Those steps include correction for bias, flat field and geometric distortion, 
and absolute flux calibration (in $W \cdot m^{-2}\cdot nm^{-1}\cdot sr^{-1}$). 
\\
The last step of the calibration transforms the images in radiance factor (hereafter noted RADF). The radiance factor was notably defined in \cite{Hapke_1993_CUP_book}, we computed it here according to the following formula:

\begin{equation}
RADF(\lambda) = \frac{\pi \cdot I(\lambda)}{F_{\odot}(\lambda, r)}
\end{equation}

where I is the observed scattered radiance, and F$_{\odot}$ the incoming 
solar irradiance at the heliocentric distance $r$ of the comet. The solar 
irradiance F$_{\odot}(\lambda, r)$ is wavelength dependent and it 
was calculated at the central wavelength of each filter to be consistent 
with the methodology applied to derive the absolute calibration factors.\\

For the spectrophotometric analysis we first coregistered the images of a given 
observing sequence using the images acquired with the F82 NAC filter (centred 
at 649.2nm) as reference. 
A python script was developed to perform the coregistration based on an 
adaptation of the python code described in \citet{VanDerWalt_2014_Peerj_Registration}.

We then generated simulated images of the solar incidence, emission and phase 
angles used for the photometric correction of the data. A 3D shape model of 
the region flown-by (hereafter Digital Terrain Model, or DTM) was created 
following the same procedure described in \cite{Preusker_2015_AA_67P_SPG}. This model 
was used alongside SPICE kernels \citep{Acton_1996_PSS_SPICE} of the 
reconstructed trajectory of the spacecraft produced by the Rosetta flight 
dynamics team of ESA and the Optimised Astrophysical Simulator for Imaging 
Systems (hereafter OASIS) \citep{Jorda_2010_SPIE_OASIS, Hasselmann_2016_Icarus_Lutetia} 
to retrieve the illumination conditions of each facet of the shape model. 

Those informations were employed to produce a photometric correction of the 
data. For the spectrophotometric analysis, we correct the images for the 
illuminations conditions by applying the Lommel-Seeliger disk law:

\begin{equation}
D(i,e,a) = \frac{2\mu_{i}}{\mu_{e}+\mu_{i}}
\end{equation}

where $\mu_{i}$ and $\mu_{e}$ are respectively the cosine of the solar 
incidence (i) and emission (e) angles.

Three-colors images (or hereafter RGB) were generated from the observations with the NAC 
filters centred at 743.7 nm (R),  649.2 nm (G), and 480.7 nm (B). These images 
were simply co-registered, no photometric corrections were applied to them. The RGBs 
were produced with the so-called STIFF program \citep{Bertin_2011_ASCL_STIFF}.
Figures \ref{fig:RGB1} and \ref{fig:RGB2} present two of the generated 
color-stretched RGB from the presented dataset, in order to investigate colour 
heterogeneities over the surface.

The spectral slope  was computed using  the following formula:

\begin{equation}
S [\%/100nm] = \frac{R_{743nm}-R_{535nm}}{R_{535nm}}\cdot \frac{10^{4}}{ (743.7 [nm] - 535.7 [nm])}
\label{Eq:Spectral slope definition}
\end{equation}

where R$_{i}$ is the radiance factor at a given wavelength. We choose to maintain 
the normalisation at 535.7 nm (F83 filter), even for sequences where we do not 
have direct measurements with the F83 filter, to be coherent with previous studies 
on the comet and literature data. Thus, for the 3 filters sequences, we estimated 
the 535.7 nm flux for each pixel through a linear regression of the data acquired 
between 480.7 nm and 649.2 nm.

\section{Global analysis of the flown-by region} \label{Section 4}
The global spectrophotometric and photometric properties of the nucleus but also
the surroundings of the flown-by region have been previously investigated by 
\cite{Fornasier_2015_AA_67PCG} (see Figs. 9, 13, and 14) on the first resolved 
images of the comet acquired on July-August 2014, at 3.6 AU of heliocentric distance. 
The surroundings of the flown-by region were classed as belonging to the group 
of terrains with the highest spectral slope (i.e. the reddest) and, in the 
882.1-535.7nm range, they were observed to have a spectral slope value > 14 \%/(100nm) 
at phase angle 1.3$^{\circ}$, and > 18\%/(100nm) at phase angle 50$^{\circ}$. 
The reader is referred to Figs. 14 and 9 in the aforementioned article for more details.

\begin{figure*}

\begin{minipage}[t]{0.41\linewidth}
\begin{center}
	\includegraphics[width=0.92\textwidth]{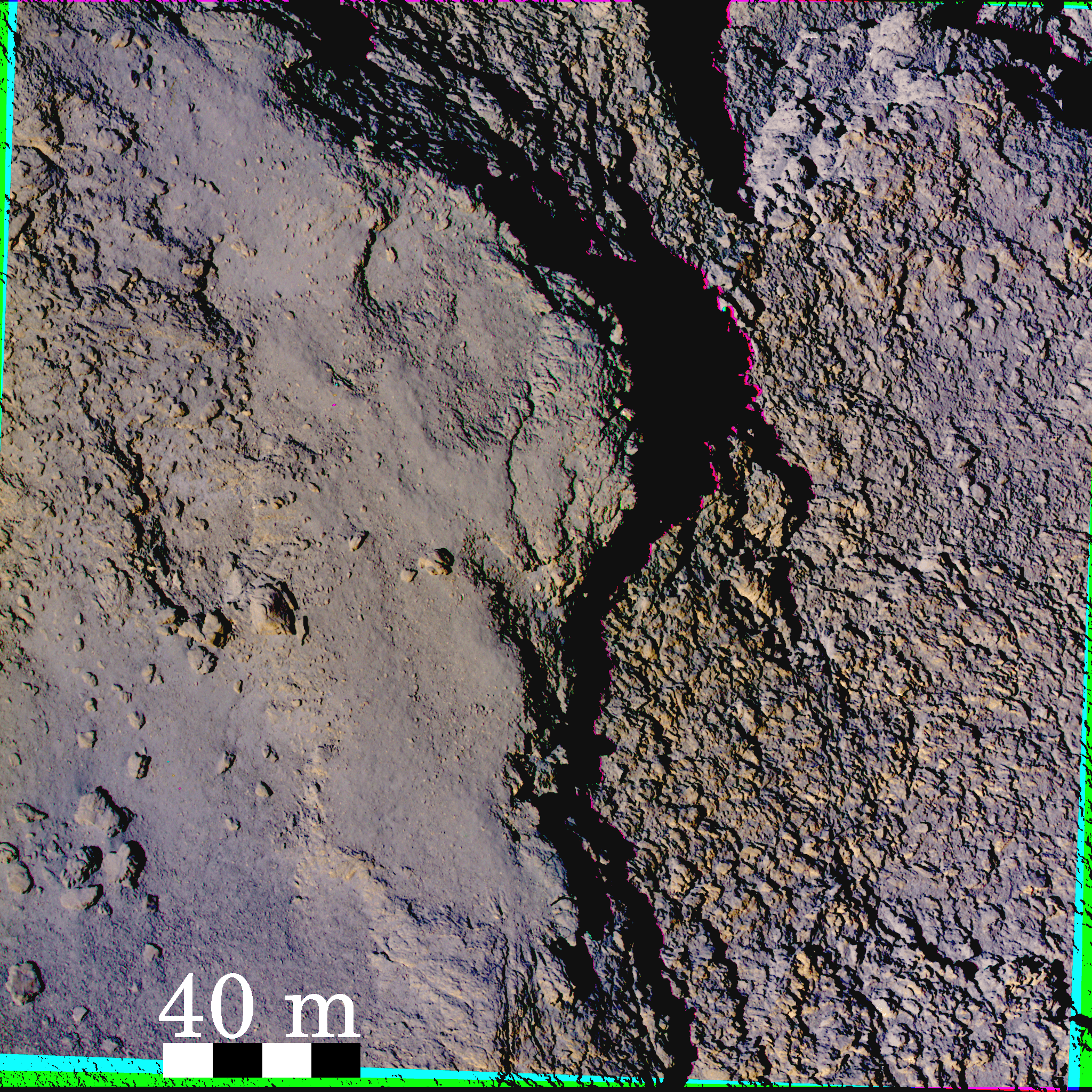}
	\subcaption{RGB map generated from the 12:20:54's sequence of observations}
	\label{fig:RGB1}

	\includegraphics[width=\textwidth]{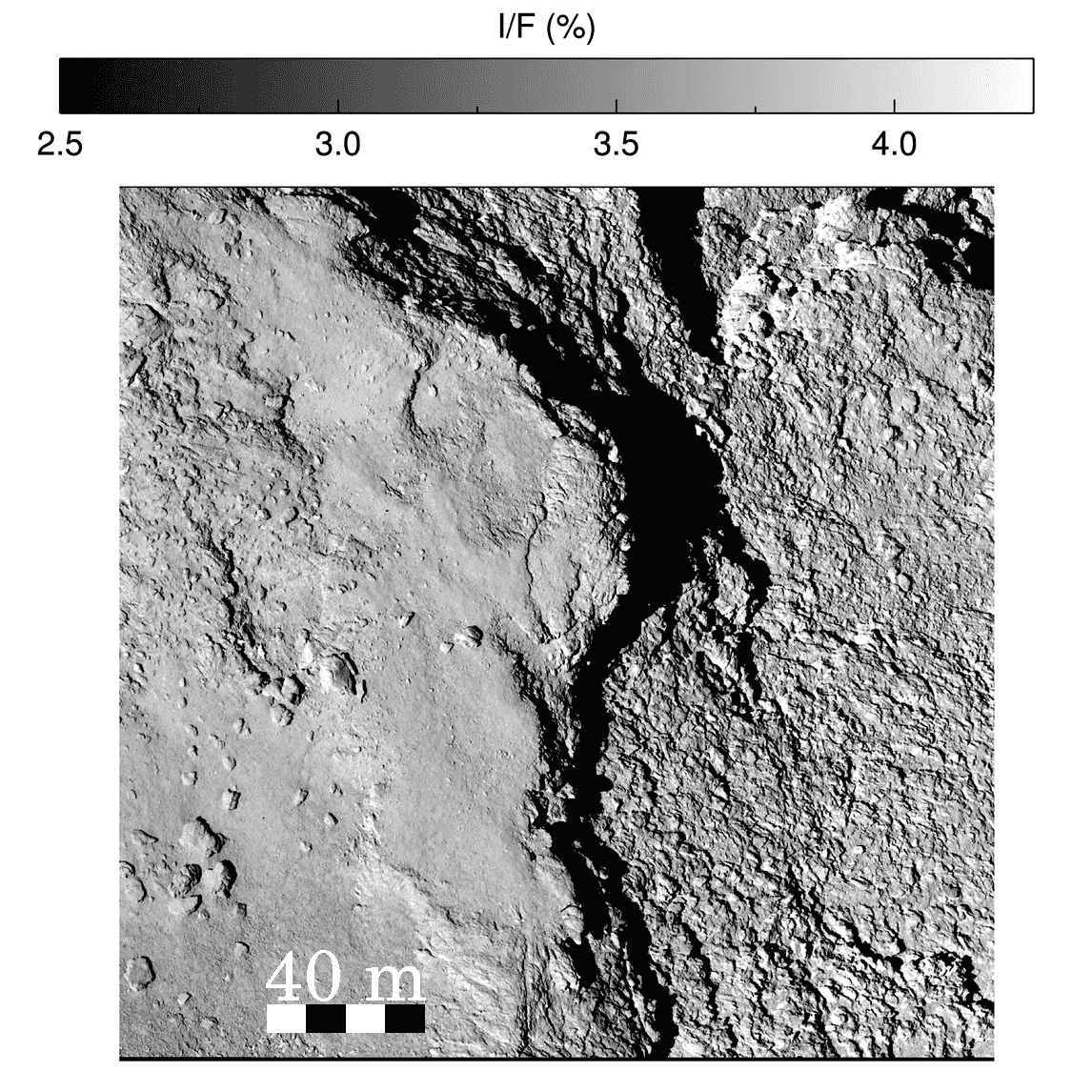}
	\subcaption{NAC Orange filter image taken at 12:20:54}
	\label{fig:Image NAC F22 ContextFrame 1}
	
	\includegraphics[width=\textwidth]{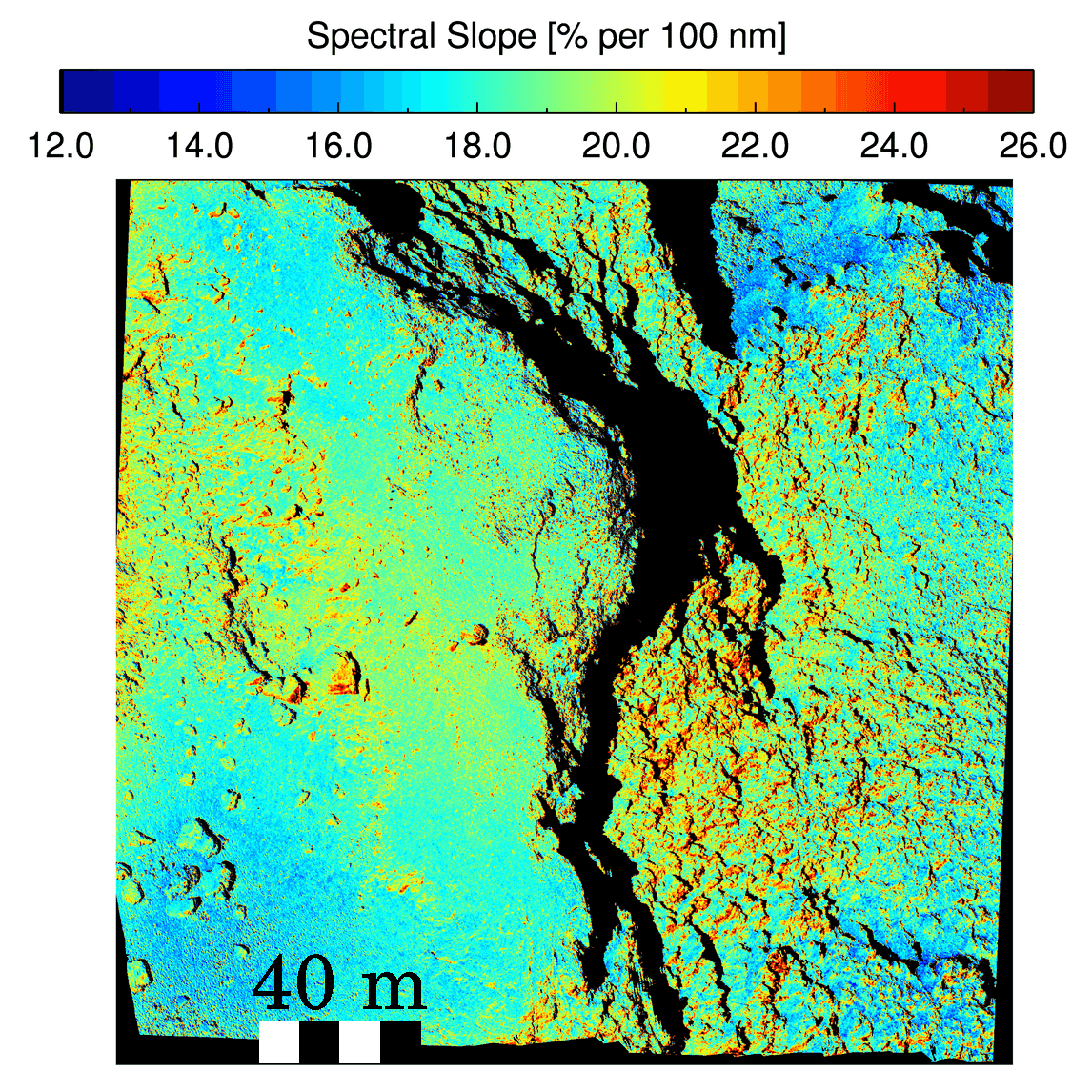}
	\subcaption{Spectral slope mapping computed from the 12:20:54's observations}
	\label{fig: Image NAC F22 12:20:54 spcslp with LS correction}
	\end{center}
\end{minipage}\hspace{1cm}
\begin{minipage}[t]{0.41\linewidth}
\begin{center}
    \includegraphics[width=0.92\textwidth]{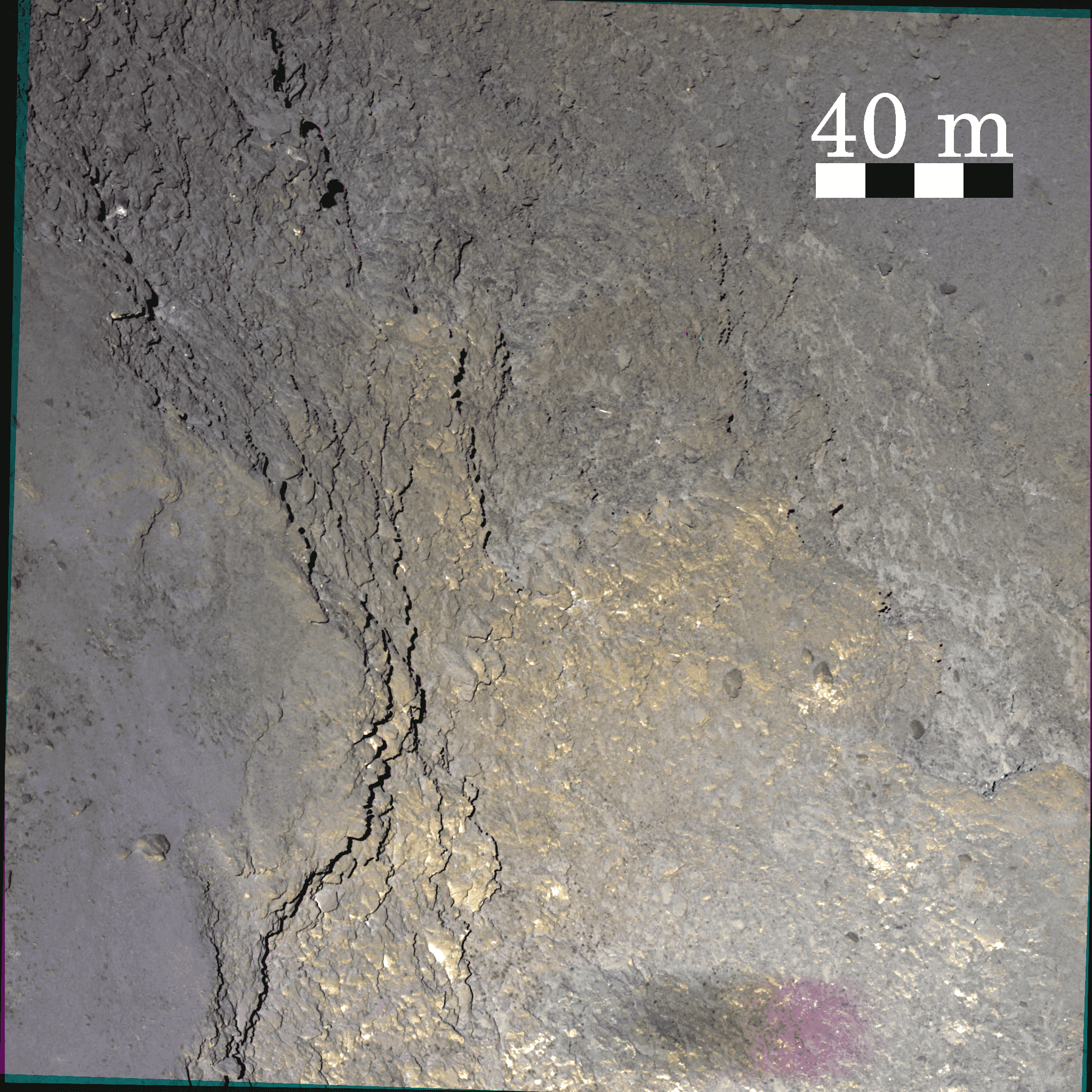}
    \subcaption{RGB map generated from the 12:39:58's sequence of observations}
    \label{fig:RGB2}

    \includegraphics[width=\textwidth]{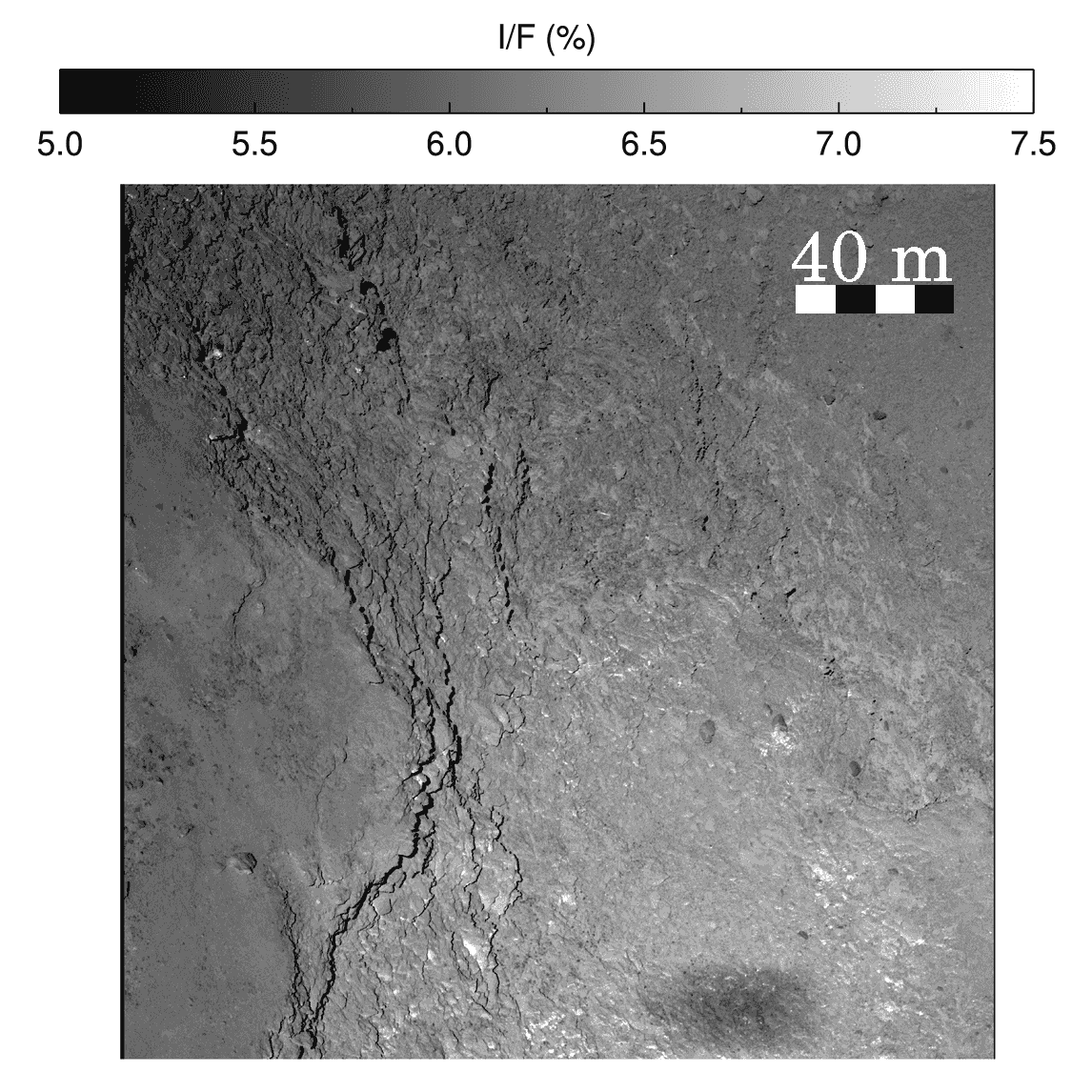}
    \subcaption{NAC Orange filter image taken at 12:39:58}
    \label{fig:Image NAC F22 ContextFrame 2}
   	
	\includegraphics[width=\textwidth]{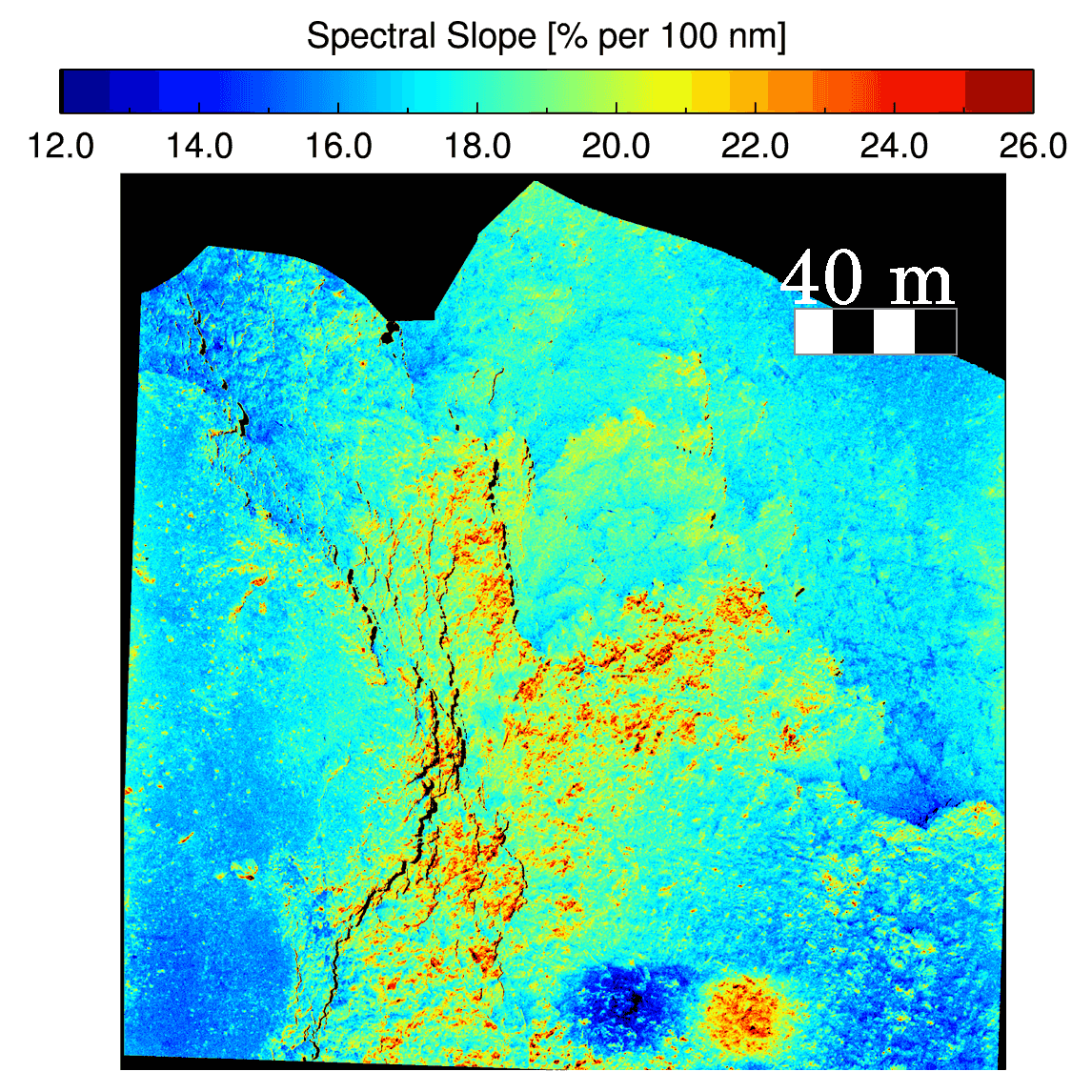}
	\subcaption{Spectral slope mapping computed from the 12:39:25's observations}
	\label{fig: Image NAC F22 12:39:25 spcslp with LS correction}
\end{center}

\end{minipage}
\caption{RGB, RADF images, and spectral slope mappings from the 12:20:54 and 12:39:25 sequences of observations}
\label{Panel1}
\end{figure*}

Fig. \ref{Panel1} depicts the RGB colours, reflectance and spectral slope maps 
of areas observed at 12:20:54 and 12:39:58. The surface clearly shows local reflectance 
and colour/spectral slope values variations, with the presence of  several bright 
spots, sombre boulders and some striae, confirming that the 
surface is heterogeneous at several scales. Brightness variations in the comet's 
surface at centimetre and millimetre scale were reported by the CIVA camera on-board the Philae 
lander \citep{Bibring_2015_Science_CIVA}, which observed a surface globally dominated by dark 
conglomerate, likely made of organics, with the presence of brighter spots that 
may be linked to mineral grains or pointing to ice-rich material.  Heterogeneities, 
reported by \cite{Fornasier_2015_AA_67PCG}, were observed at a larger scale (> 2 m/pxl).

From the 14$^{th}$ February images we find a median reflectance of 6.15$\pm$
0.07\% (cf. Fig. \ref{fig:Image NAC F22 ContextFrame 2}) in the F82 filter image taken at 12:39:58.  The sombre 
boulders show a reflectance $\sim$ 10\% lower than the average value while 
the brightest regions show a reflectance about 20\% higher. However, in this 
flown-by region, we do not see strong variations (i.e. a factor 2 or higher) in 
reflectance as observed for bright spots and ice rich regions on different areas 
of the nucleus 
\citep{Pommerol_2015_AA_ExposedIce, Barucci_2016_AA_Ices_Features,Oklay_2016_AA_Variegation,Fornasier_2016_Science_IceDust}. \\
The most remarkable changes are observed in the images taken when the phase angle 
is minimal, such as in Fig. \ref{fig:Image NAC F22 ContextFrame 2}.
In particular, we notice several boulders, metre-sized and sub-metre-sized,  located on both 
sides of the region, whose reflectance is lower than that of their surroundings at small phase 
angle observations. We determined that the phase functions of those boulders are different from 
that of observed bright spots, or even the immediate surroundings of such features. 
This particular result is discussed in the next section alongside the photometric analysis of 
the entire region (cf. Fig. \ref{fig: Phase functions} and table \ref{Table:HHS - Fast Hapke model}).

The RGB images (see figs. \ref{fig:RGB1} and \ref{fig:RGB2}) indicate redder colours along the cliff in Fig. \ref{fig:RGB1}, and
along the cliff, terraces, and outcrops in Fig. \ref{fig:RGB2}.  In this image, 
the green to red artefact in the bottom is simply the shadow projected by the Rosetta spacecraft. 
Similarly, the receding of the projected shadows and the motion of the S/C also left 
artefacts in the RGB. Such elements on surface of the nucleus were not further considered.

In those colour-stretched RGBs, we note that most of the previously spotted bright patches and 
striae of surface appear to be located where the nucleus surface looks redder. This is confirmed 
by the spectral slope mappings (in the 743-535 nm range, Figs. \ref{fig: Image NAC F22 12:20:54 spcslp with LS correction} 
and \ref{fig: Image NAC F22 12:39:25 spcslp with LS correction}), where the aforementioned 
red features display indeed a strong spectral slope (above 21\%/100nm). Spectral slopes values 
are found to vary between 15 and 23 \%/(100nm)  at phase angle $\sim$ 1$^{\circ}$  (UTC: 
12:39:58, cf. Fig. \ref{panel: spectral slopes}), confirming that the flown-by region 
is one of the redder of the comet's surface.  \\
The spectral slope mappings evolution computed at different epochs of the flyby 
is represented in Fig. \ref{panel: spectral slopes} and the corresponding phase 
reddening effect is discussed in section \ref{par:phase reddenning} and summarised 
by Fig. \ref{fig:spectral slope vs phase angle}.

\begin{figure*}
\includegraphics[scale=0.25]{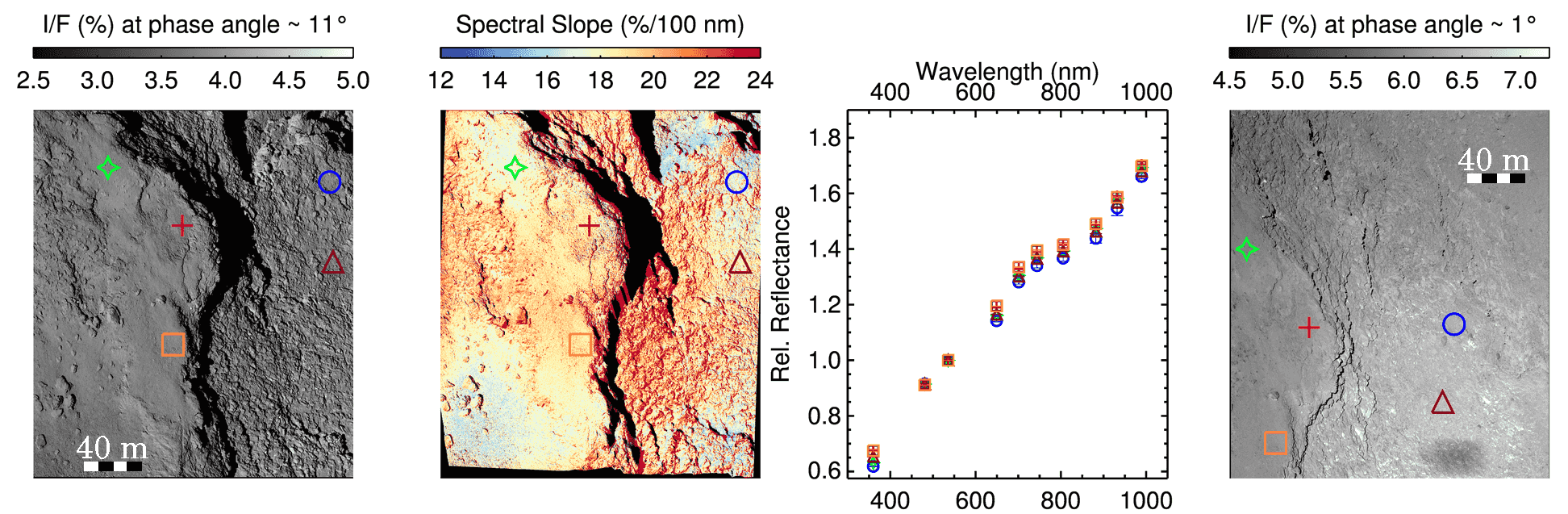}
	\caption{Radiance factor of the flown-by region, associated spectral slope map 
(evaluated in the 743-535 nm range) and spectrophotometry for 6 selected features 
at phase angle 11 $^{\circ}$  (NAC F22 image UTC - 12:20:54). On the left, the 
radiance factor at phase $\sim$ 1 $^{\circ}$ (NAC F22 image UTC - 12:39:54) is also represented.}
	\label{panel: IDAB-12h19}
\end{figure*}

We investigated the spectrophotometric properties of the surface on several 
relatively smooth areas visible on multiple sequences and on both morphological 
regions, and where no boulders are discernible at the decimetre scale. We considered 
the observations related to the 12:20:54 and 12:39:58 sequences.  
The integrated radiance factor inside boxes of 15x15 pixels (i.e. about 1.5x1.5m$^{2}$ 
on the nucleus) is presented in Fig. \ref{panel: IDAB-12h19}. We note that, at every 
wavelength, the variations of the measured relative reflectances are, in average, 
of 3.5$\pm$2.1\%. Furthermore, in the 743-535.5nm range, the median spectral 
slope of those measurements is found to be equal to (17.7$\pm$1.0)\%/100nm.
Hence, the smooth terrains investigated over the two different morphological regions 
were not found to be distinctly different.

We also note an increase of the flux in the 700-750 nm region, as already seen 
in \cite{Fornasier_2015_AA_67PCG} analysis. These feature is not related to the nucleus 
surface properties but it has been linked to cometary activity and attributed 
to  H$_{2}$O$^{+}$ and/or NH$_{2}$ emission in the very inner coma 
\citep{Fornasier_2015_AA_67PCG}. As our observations are mostly carried out in 3 filters, 
including the one centred to 743 nm, we need to estimate the cometary emission 
effects in the spectral slope value  evaluated in the 743--535 nm range compared 
to the one evaluated in the 882--535 nm wavelength range, and used in the 
literature for the 67P/CG comet \citep{LaForgia_2015_AA_Agilkia, Fornasier_2015_AA_67PCG, Oklay_2016_AA_Variegation, Pajola_2016_AA_67P_Aswan_review}. 
We thus computed the spectral slopes for those investigated terrains, in both 
those wavelength ranges, for the observations acquired at 12:09:54 and 12:20:54, 
where the full set of NAC filters is available.
We found that the spectral slopes computed in the 743--535 nm range are about 
24\% higher that those computed in the  882--535 nm range, thus indicating an 
important contribution of the cometary emission. The different spectral wavelength 
range used and the cometary emission contribution in the 743-535 nm spectral slope 
must be taken into account when comparing spectral slopes values reported 
in the literature.

When considering the reflectance maps of figures \ref{fig:Image NAC F22 ContextFrame 1}
and \ref{fig:Image NAC F22 ContextFrame 2} more closely, we observe numerous local 
heterogeneities in terms of absolute reflectance and of spectral slope value, that we discuss in 
the following section. \\

\section{Reflectance and spectral behaviour of local features}
\begin{figure*}
\centering
	\includegraphics[scale=0.347]{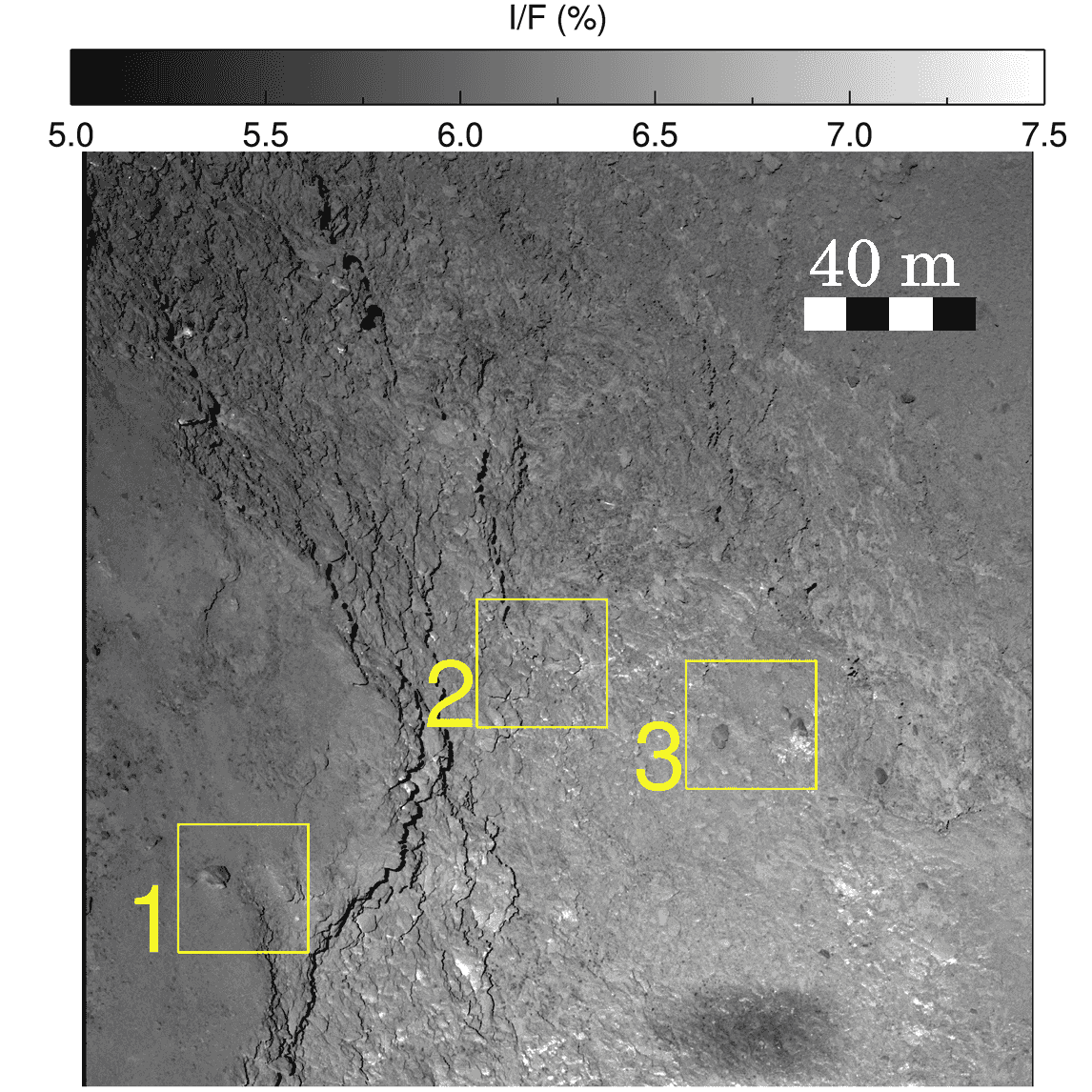}
	\caption{NAC orange filter image taken at UTC - 12:39:58 with the relative locations of Regions of Interests.}
	\label{Fig: 12h38b 3ROIs}
\end{figure*}

\begin{figure*}
	\includegraphics[scale=0.25]{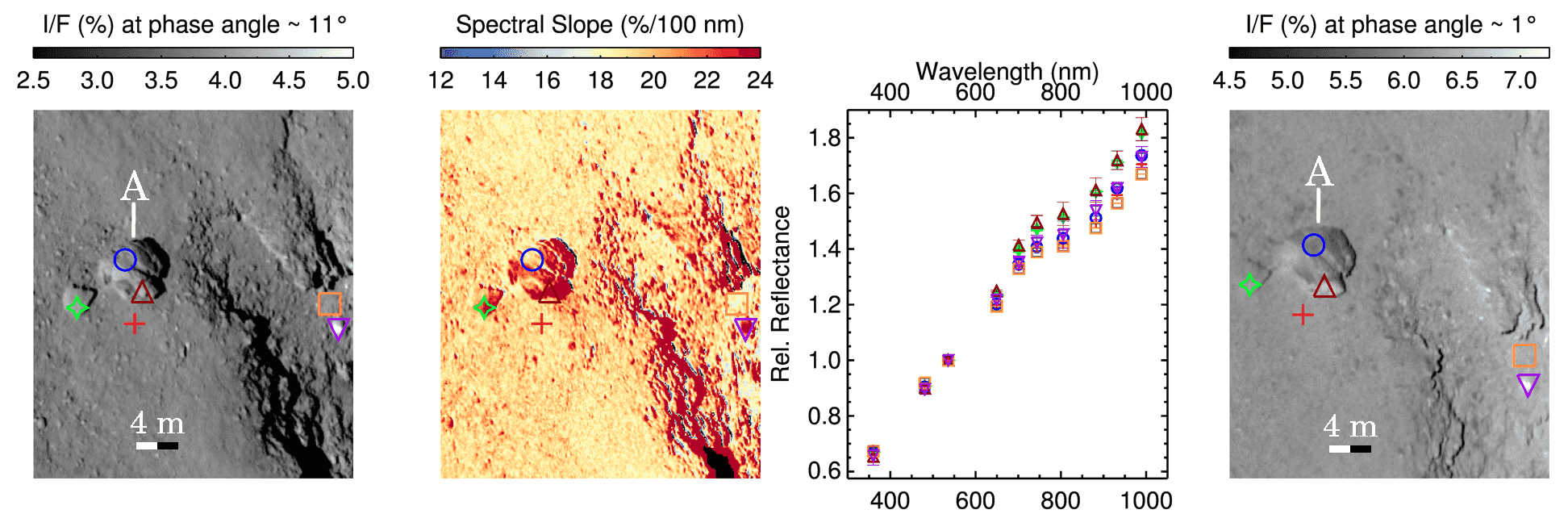}
	\caption{From left to right: Radiance factor of ROI 1, associated spectral slope 
map (evaluated in the 743-535nm range) and spectrophotometry for 6 selected features 
at phase angle 11$^{\circ}$ (NAC orange filter UTC - 12:20:54). Radiance factor 
at phase $\sim$ 1$^{\circ}$ (NAC orange filter UTC - 12:39:54).}
	\label{panel: ID51-12h19}
\end{figure*}

\begin{figure*}
	\includegraphics[scale=0.25]{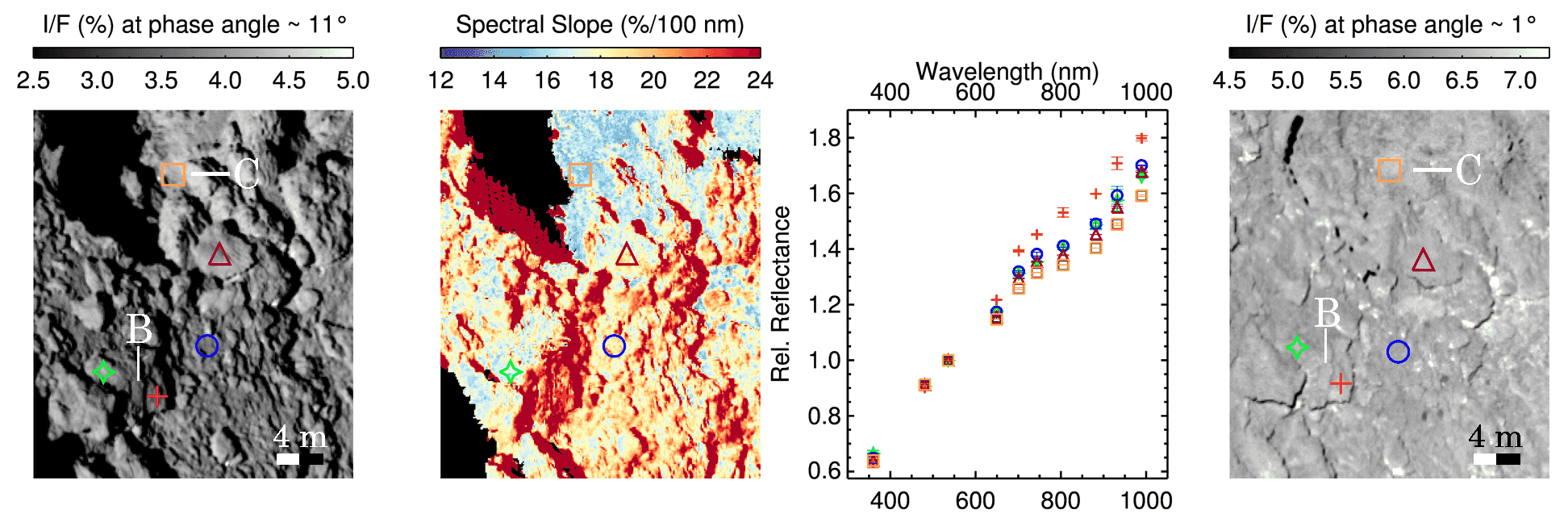}
	\caption{From left to right: Radiance factor of ROI 2, associated spectral slope 
map (evaluated in the 743-535nm range) and spectrophotometry for 5 selected features 
at phase angle 11$^{\circ}$ (NAC orange filter UTC - 12:20:54). Radiance factor 
at phase $\sim$ 1$^{\circ}$ (NAC orange filter UTC - 12:39:54).}
	\label{panel: ID59-12h19}
\end{figure*}

\begin{figure*}
	\includegraphics[scale=0.25]{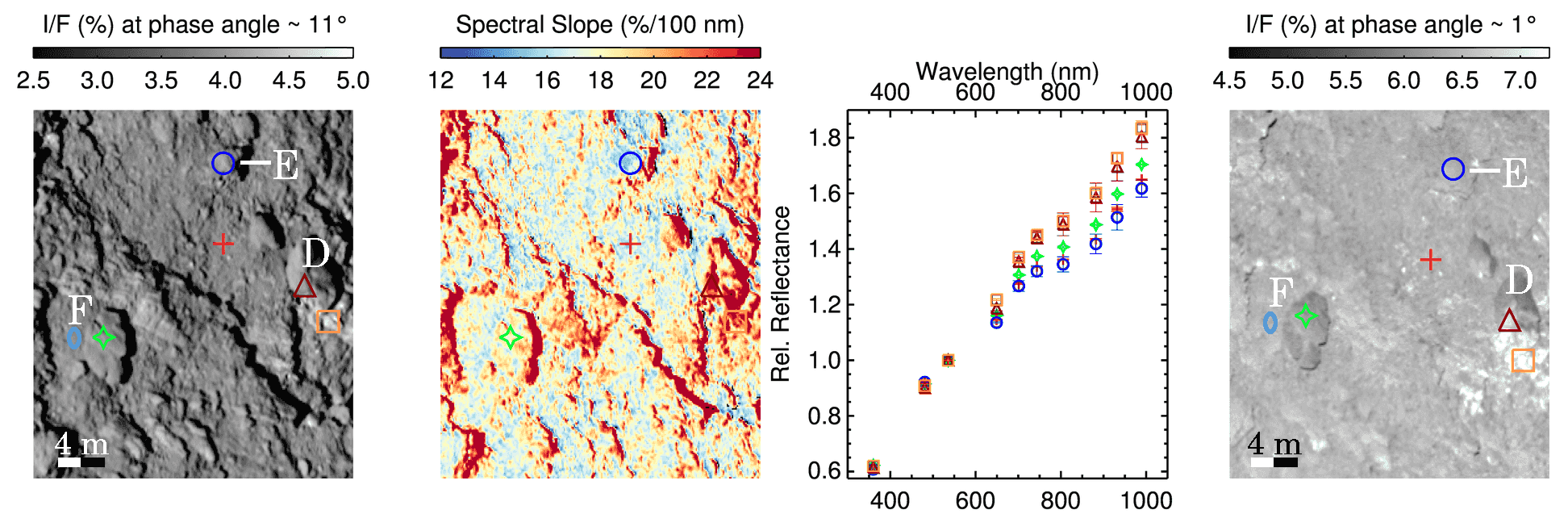}
	\caption{From left to right: Radiance factor of ROI 3, associated spectral slope 
map (evaluated in the 743-535nm range) and spectrophotometry for 5 selected features 
at phase angle 11$^{\circ}$ (NAC orange filter UTC - 12:20:54). Radiance factor 
at phase $\sim$ 1$^{\circ}$ (NAC orange filter UTC - 12:39:54).  The blue patch under the letter F refers to the position of the measurement at the base of the boulder in Fig.\ref{fig: Phase functions}.}
	\label{panel: ID65-12h19}
\end{figure*}

For this analysis, we sought surface elements, which were visible in almost every 
single sequence of observations, and which also exhibited heterogeneities in terms 
of absolute reflectance and of spectral slopes. We investigate here 3 such regions 
of interests (ROI), which are indicated in Fig. \ref{Fig: 12h38b 3ROIs}.
The first ROI is over the Ash region and includes two metre-scale boulders as 
well as a spot ( the violet triangle in Fig. \ref{panel: ID51-12h19}) which appears 
bright on every filter image of the dataset.
The second ROI delimits a part of the Imhotep region including a strata head which 
appears dark and present a red spectral behaviour (site B in Fig. \ref{panel: ID59-12h19})
 as well as an outcrop exhibiting a blue spectral behaviour 
(site C in the same figure). Finally, in the third ROI, we investigated a couple 
of boulders which clearly appear darker than the surroundings at small phase angle 
(sites D and F in Fig. \ref{panel: ID65-12h19}).

More in detail, ROI no. 1 of Fig. \ref{Fig: 12h38b 3ROIs} is the clear example of the fine deposits 
on the layered terrain, this being demonstrated by the eroded niche margin on 
the right hand side, unveiling a set of cm-scale layers. ROI no. 2 focuses on 
the decimetre-scale layers with the possible presence of detaching blocks, as 
observed in Hathor region, but on different scales \citep{Pajola_2015_AA_BoulderSizeDistribution}. 
Finally, ROI no. 3, corresponds to a specific area with a cluster of boulders. The origin 
of this kind of boulders has not been here investigated, and it is not the purpose 
of this work, nonetheless, since there a cliff in close proximity it is not unreasonable 
to think that they may have detached from it, as observed and confirmed on many 
other locations on 67P/CG \citep{Pajola_2016_AA_67P_Aswan_review}.

In those ROIs, the spectrophotometric properties of different features was 
investigated by integrating the radiance factor, on Lommel-Seelinger disk-corrected 
images, inside squared boxes of 3x3 pixels (i.e. about 0.33x0.33m$^{2}$).

The spectrophotometric measurements in ROI 1 (cf. Fig. \ref{panel: 
ID51-12h19}) indicate that boulder named A is very red in term of spectral slope ($S$ = 22.5 
\%/(100nm)).  Its brighter side, indicated by the violet 
triangle, is  spectrally redder that the darker regions of the same boulder, indicated by the blue 
circle, and as red as the nearby boulder on its left (green star). The bright spot indicates by the 
brown triangle has a spectral slope less steep than that of boulder A ( $S$ = 20 
\%/(100nm)), but it is still redder compared to the relatively smooth regions 
indicated by the red cross and orange square, whose spectral slope value is around 19
\%/(100nm).
The behaviour of this surface element, i.e. high reflectance and also steep spectral slope, is 
completely different compared to what previously observed by 
\cite{Fornasier_2015_AA_67PCG, Pommerol_2015_AA_ExposedIce, Fornasier_2016_Science_IceDust, Barucci_2016_AA_Ices_Features,Oklay_2016_AA_Variegation}. 
In those papers, the authors observed bright features on Hapi and in other 
locations over the nucleus, which were associated with a small spectral slope 
value ($S$ < 10 \%/(100nm)). This behaviour, i.e. high reflectance and 
moderate spectral slope value, was linked to the presence of water-ice mixed to 
refractory material. 
We surmise that the red spectral slope of the bright feature in ROI 1 could be explained
through a difference in terms of composition: this feature would not be 
enriched in water-ice but probably rather in minerals brighter than the 
surroundings dark agglomerate. But it could also point out to a difference of 
texture: this feature might be smoother or have a lower porosity than its 
surroundings.\\
Additionally in this ROI, we observe that the top of A boulder (blue circle) 
has a spectral behaviour similar to that of the surface layer at its feet (red 
cross) indicating that both are covered by similar dark and red dust material.

The area contained in ROI 2 (cf. Fig.\ref{panel: ID59-12h19}) displays higher variations in the 
spectral behaviour compared to ROI 1, including both red features such as the one 
named B ($S$ = 21.7 \%/(100mn)), and bluer regions such as the one named C 
($S$ = 15\%/(100mn)). 
Strata-head B is characterised by a steep spectral slope in the 650-1000 nm range (see the red cross 
symbol on Fig. \ref{panel: ID59-12h19}) among the investigated areas. The measurement 
on the outcrop named C (where the orange square is placed) has a slightly higher 
radiance factor than the average in ROI 2 together with a bluer spectral behaviour 
than the surroundings.

The two metre-sized boulders located beneath the green cross and brown triangle 
have a spectral behaviour indistinguishable from the smoother region indicated 
by the blue circle, again pointing towards a similar composition in the uppermost dust layer.

In ROI 3, we have investigated two metre-scaled boulders which show again a lower 
opposition effect at small phase angles, designated D and F in Fig. \ref{panel: ID65-12h19}
together with another boulder, named E, whose reflectance is indistinguishable from the 
surroundings at small phase angle. This region shows spectral slope variations as high as in ROI 2. 
Particularly interesting is feature D, which entails some bright patches at the feet of the boulder. 
The reflectance of those patches is up to 40\% higher than the boulder's. Despite the reflectance 
variation, both dark and bright components of boulder D show a  similar very red spectral behaviour, 
redder than smooth surrounding regions (as the one indicated by the red cross) or of F and E boulders.

Our interpretation is that, possibly due to thermal stress, the sombre boulder has fractured and 
released chunks of fresh material that now constitute the bright patches. Both bright patches 
and boulder might have been later coated with organic-laced dust, and therefore exhibit such a 
red spectral slope (see Fig. \ref{fig:RGB2}).

\subsection{Phase reddening}\label{par:phase reddenning}
We plotted the spectral slope values versus the phase angle for each pixel of the available 
images is represented in Fig. \ref{fig:spectral slope vs phase angle}. 

\begin{figure*}
\centering
	\includegraphics[scale=0.30]{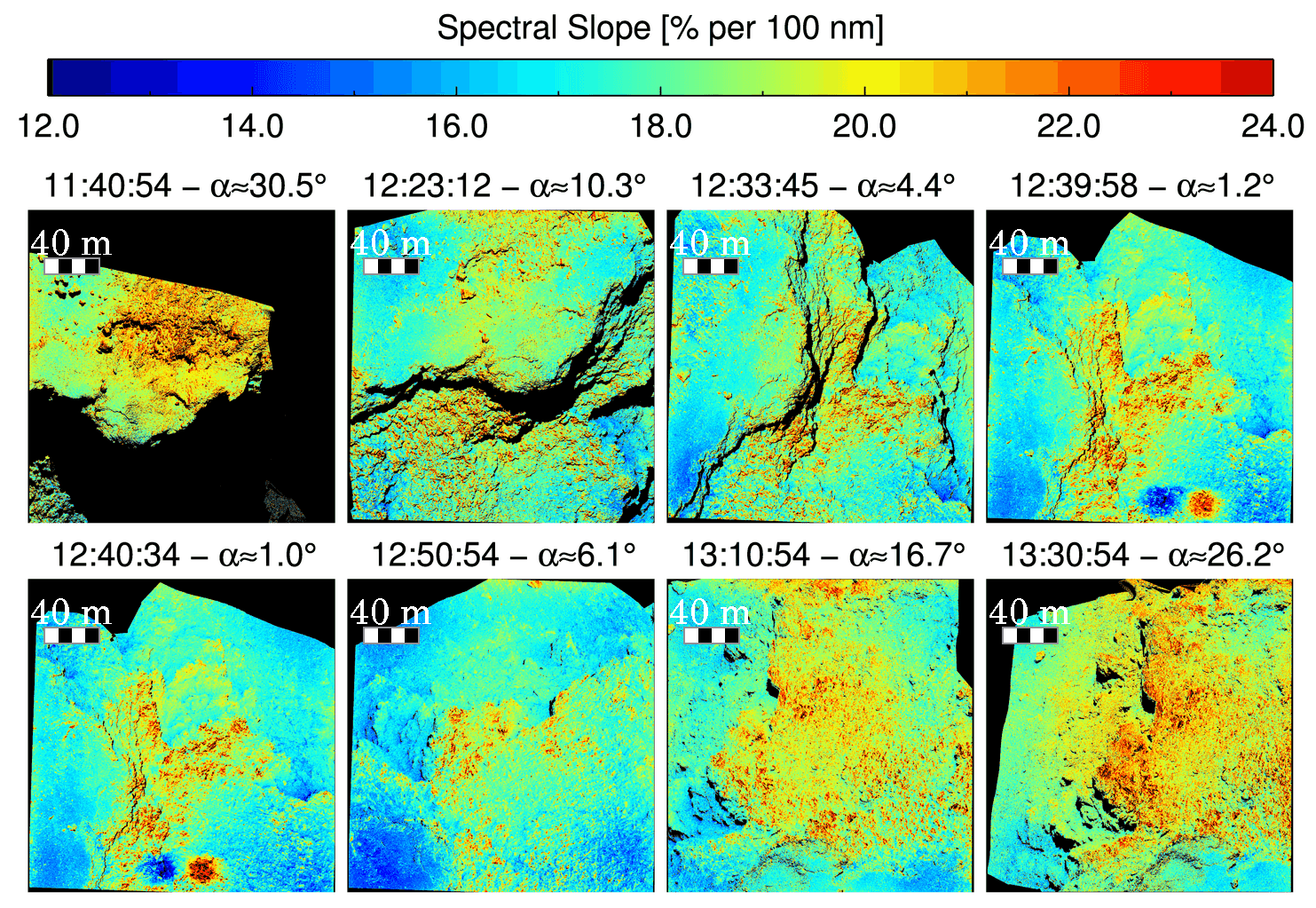}
	\caption{Panel of spectral slope mappings in the 743-535nm range at selected times of the flyby.}
	 \label{panel: spectral slopes}
\end{figure*}

\begin{figure}
\centering
	\includegraphics[width=0.49\textwidth]{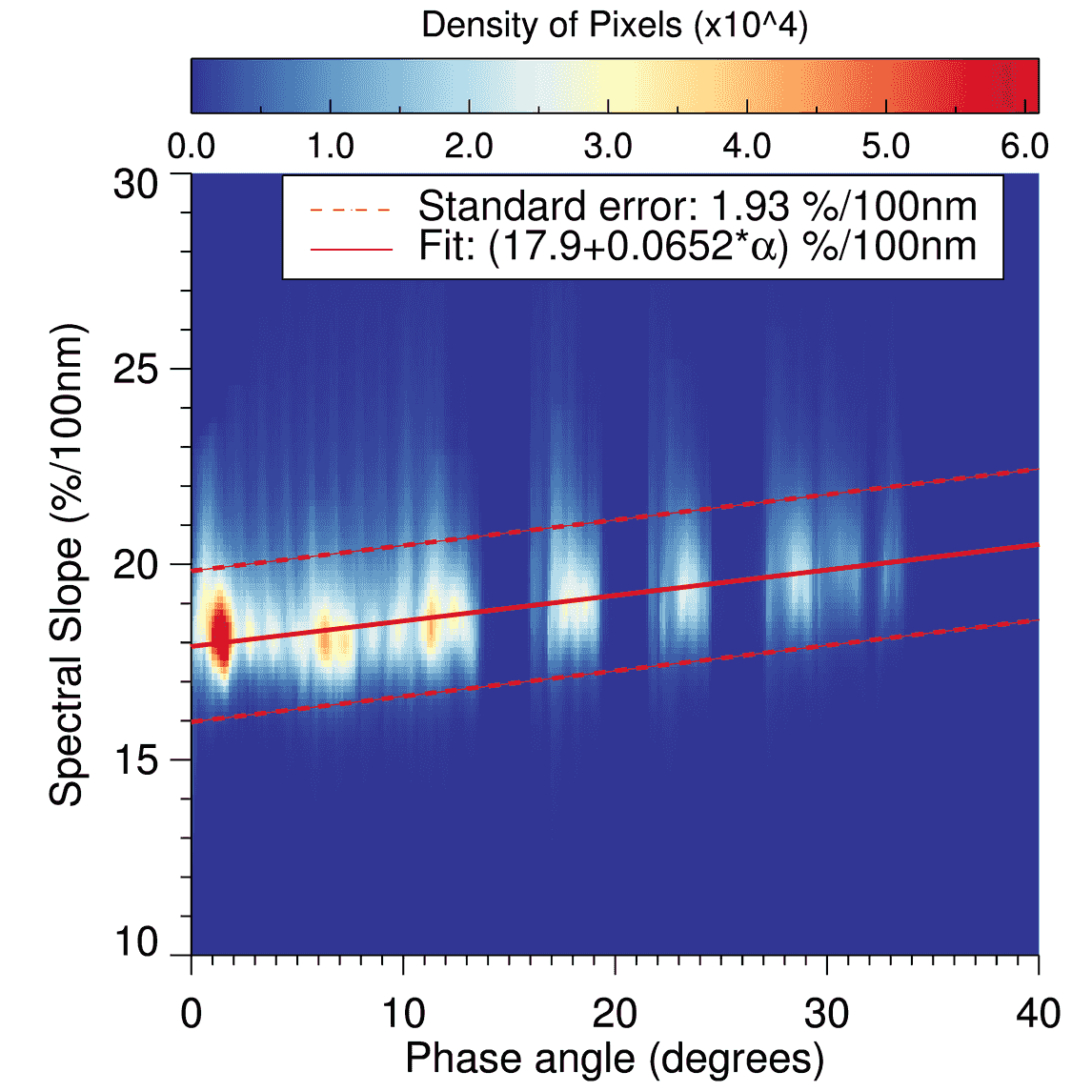}
	\caption{Spectral slope values versus phase angle. The red line corresponds 
to the linear regression of the spectral slope as a function of phase angle, while 
the dashed lines corresponds to the standard error of this fit (S.E. = 1.93191).}
	\label{fig:spectral slope vs phase angle}
\end{figure}

The phase reddening phenomenon is evident but characterized by a lower slope 
compared to the previous measurements \citep{Fornasier_2015_AA_67PCG}: a linear 
regression of this dataset gave us a slope of $\beta$ =   ($0.0652\pm0.0001$)$\cdot 10^{-04}$ 
nm$^{-1}/^{\circ}$ and the spectral slope at zero phase angle is estimated  to be 
(17.9$\pm0.1) \%/(100 nm)$ in the 743--535 nm range.

We note that \citet{Fornasier_2016_Science_IceDust} also reported a strong decrease of the 
phase reddening effect comparing the spectral slope values averaged over the comet rotational 
period from Aug. 2014 datasets ($\beta = 0.104 \cdot 10^{-04} nm^{-1}/^{\circ}$) 
and those acquired on April-Aug 2015 ($\beta=0.048 \cdot 10^{-04} nm^{-1}/^{\circ}$), 
approaching the perihelion. They explain the decrease of the phase reddening effect as a 
results of the increasing level of activity that progressively removed the dust mantle. 
The fly-by data have intermediate phase reddening value comparable to those analysed 
in \citet{Fornasier_2016_Science_IceDust}. This can be interpreted as progressive dust 
mantle removal with changes of the surface properties, like the dust roughness. 

\section{Photometric analysis}
In order to compare our results with the literature, we choose to apply the Hapke radiative transfer 
model to fit our dataset for the purpose of this study.
\subsection{Hapke model and inversion procedure} \label{Subsection:  HHS reference}
The latest equation of Hapke model, in radiance factor, is defined by the expression:

\begin{equation}
\begin{array}{c}
RADF(\lambda,\mu_{0},\mu,\alpha) = K\frac{w_{\lambda}}{4}\left(\frac{\mu_{0e}}{\mu_{e}+\mu_{0e}}\right)\times[(1+B_{SH(\alpha)})p_{(\alpha)}\\
+(1+B_{CB(\mu_{0},\mu,\alpha)})(H_{(\mu_{0}/K,w_{\lambda})}H_{(\mu/K,w_{\lambda})}-1)]\times S_{(\mu_{0},\mu,\alpha)}
\end{array}
\end{equation}

where $\mu_{0e}$ and $\mu_{e}$ are the effective cosine of incidence
and emergence angles, involving the topographic correction of the
facet by the macro-roughness shadowing function $S$. \\
$K$ is the porosity factor, $w_{\lambda}$ is the single-scattering albedo (or
SSA), $B_{SH}$ is the shadow-hiding opposition effect (SHOE) term,
$p$ is the single or double lobe Heyney-Greenstein particle phase
function (or SPPF), $B_{CB}$ is the coherent-backscattering opposition
effect (CBOE, \citealp{1988JPhys..49...77A}) term and $H$ is the
isotropic multiple scattering approximation function (also noted IMSA), 
analytically described by the second-order approximative
Ambartsumian-Chandrasekhar function \citep{2002Icar..157..523H}.
Table \ref{tab:hapke-description} presents the free Hapke parameters
and their respective relationship with the scattering curve morphology
and properties of the surface. We name this version as Helfenstein-Hapke-Shkuratov
model (hereafter noted HHS model). A further qualitative and mathematical description
of the photometric model and its functions can be found in \citet{2011Icar..215...83H}
or \citet{9781139025683} .

For data inversion we proceed as follow: first, a table of facets for each image 
is binned into a $50\times50\times50$ ($125.000$) cell grid (in steps of  
$1.5^{\circ}\times1.5^{\circ}\times(0.05^{\circ})$, which translates in to shells of 0.3 m in diameter) in respect to 
$i$ (incidence), $e$ (emergence) and $\alpha$ (phase) angles. The RADF is 
averaged for each cell, mitigating any effect of variegation and poor pixels/facets. 
Cells with no pixels are removed from data. Then, the fitting procedure is 
focused on minimizing the $\chi^{2}$ between the measured and modelled 
RADF weighted by the number of facets on each active cell. 30 initial conditions 
of lowest $\chi^{2}$ are selected from 1000 random initial conditions (search 
grid of 0.01 interval) to undertake thorough minimization by the 
Broyden-Fletcher-Goldfarb-Shanno (L-BFGS-B) algorithm 
\citep{BROYDEN01031970, Zhu:1997:ALF:279232.279236}, available by Scipy 
Python package \citep{Scipy}. L-BFGS-B solves non-linear problems by 
approximating the first and second derivatives in an iterative procedure to 
search for local solution. Finally, the 30 Hapke solutions are averaged to 
obtain the final solution and the uncertainties tied to each Hapke parameter. 

\begin{center}
\begin{table*}
\centering{}%
\begin{tabular}{c>{\centering}p{8cm}>{\centering}p{1.7cm}}
{\footnotesize{}parameter} & {\footnotesize{}Description} & {\footnotesize{}Bounds}\tabularnewline
\hline 
{\footnotesize{}$w_{\lambda}$} & {\footnotesize{}Particle Single-Scattering Albedo (SSA).} & {\footnotesize{}$\left\{ 0.01,0.5\right\} $}\tabularnewline
\hline 
{\footnotesize{}$g_{sca,\lambda}$} & {\footnotesize{}Asymmetric factor. Coefficient of the mono-lobe Heyney-Greenstein
function. The average cosine of emergence angle of the single particle
phase function (SPPF).} & {\footnotesize{}$\left\{ -1.0,1.0\right\} $}\tabularnewline
\hline 
{\footnotesize{}$K$} & {\footnotesize{}Porosity factor $K=\ln(1-1.209\phi^{2/3})/1.209\phi^{2/3}$,
$\phi$ is the filling factor. It is an addition introduced by Hapke (2008) corresponding 
to the role of superficial porosity in the scattering. The porosity factor and $h_{s}$ are 
constrained through the approximative formula $K=1.069+2.109h_{s}+0.577h_{s}^{2}-0.062h_{2}^{2}$ } 
& {\footnotesize{}$\left\{ 1.0,1.7\right\} $}\tabularnewline
\hline 
{\footnotesize{}$Bs_{0}$,$h_{s}$} & {\footnotesize{}Amplitude and angular width of the SHOE. $h_{s}$
relates with micro-roughness and porosity. $Bs_{0}$is a function
of the specular component of the Fresnel reflectance $S(0)$.} & {\footnotesize{}$\left\{ 0.0,3.0\right\} $,}{\scriptsize \par}
{\footnotesize{}$\left\{ 0.0,0.15\right\} $}\tabularnewline
\hline 
{\footnotesize{}$Bc_{0}$,$h_{c}$} & {\footnotesize{}Amplitude and angular width of the CBOE. $h_{c}=\frac{\lambda}{2\pi l}$,
thus $h_{c}$ is inversely proportional to the mean photon path $l$.
$Bc_{0}$ is somehow connect to the particle scattering matrix.} & {\footnotesize{}$\left\{ 0.0,1.0\right\} $,}{\scriptsize \par}
{\footnotesize{}$\left\{ 0.0,0.15\right\} $}\tabularnewline
\hline 
{\footnotesize{}$\bar{\theta}$} & {\footnotesize{}Average macroscopic roughness slope of sub-pixel/sub-facet
scale.} & {\footnotesize{}$\left\{ 1^{\circ},90^{\circ}\right\} $}\tabularnewline
\hline 
\end{tabular} 
\protect\caption{\label{tab:hapke-description} Description of the Hapke Parameters.}
\end{table*}
\par\end{center}

The photometric phase curve of the area, sampled between $0.2^{\circ}$ and $33^{\circ}$, 
shows a similar behaviour in the 3 filters investigated (NAC F82, NAC F84 and NAC F88): 
linear up to $\alpha\approx5^{\circ}$, slightly non-linear up to $\alpha=0.2^{\circ}$ and 
without any obvious sign of sharp spike at close opposition. We considered all the pixels
covered by the shape model at $i$ and $e$ smaller than $80^{\circ}$. We also further 
excluded those affected by the spacecraft shadow, i.e. $\alpha\leq 0.2$ . Thus, we 
used a total of 25 NAC F82, 24 NAC F84 and 23 NAC F88 images, which amounts to
respectively, a total of 650,810, 633,674 and 647,776 active cells, reaching phase 
angles low enough to constrain the CBOE parameters. Most of the RADF spread is 
accounted by the assorted illumination conditions, while some few scattered cells 
are due to the poor description of the shape model for certain terrain slopes or 
overshadowed facets. 

The HHS model has a maximum of 10 free parameters to be adjusted when the photometric 
sampling has a large enough coverage, i.e., from $0^{\circ}$ to $\sim90^{\circ}$ for the $i$ and $e$ 
angles, and $\alpha$ from $0^{\circ}$to $\sim180^{\circ}$. However, the flyby data are limited
to $\alpha\leq 33^{\circ}$, leading us to apply only the single-term Heyney-Greenstein 
function to describe the backscattering lobe of the particle phase function. By also taking into 
account the approximative expression between $h_{s}$ and $K$ parameters (Table \ref{tab:Hapke-parameters}, 
\citealp{2011Icar..215...83H}), we end up with 7 free parameters.

\subsubsection{Role of the CBOE term}
The coherent-backscattering mechanism rises from inter-intra-grain multiple scattering 
of secondary and higher orders. During our modelling, the first iteration of the inversion 
procedure revealed that CBOE parameters do not gradually converge on \citet{2002Icar..157..523H} 
or HHS models to a single solution, which points out to redundancy of this mathematical 
addendum when a second opposition spike is not verified. This have also been emphasized 
in our analysis of the first approaching images of the whole northern hemisphere 
\citep{Fornasier_2015_AA_67PCG}.  Thus, visual inspection of phase curve have shown 
no presence of any sharp opposition spike to $\alpha\leq0.2$. \citet{2007JGRE..112.3001S} 
have also found out CBOE is negligible when analysing samples with \citet{2002Icar..157..523H} 
model. Therefore, considering that coherent-backscattering is not expected to play a major 
role in the opposition effect of high absorbance surfaces \citep{2010epsc.conf..738S,2012Icar..217..202S}, 
such as the cometary carbon-rich mantles, we decided to remove it from our final Hapke analysis. 
We note that also \cite{Masoumzadeh_2016_ESLAB} found the coherent back-scattering 
mechanism negligible in their analysis of the 14$^{th}$ Feb. 2015 data of the 67P/CG comet.

\subsection{Results}
\begin{figure*}

\begin{minipage}[b]{0.49\linewidth}
    \centering
	\includegraphics[scale=0.051]{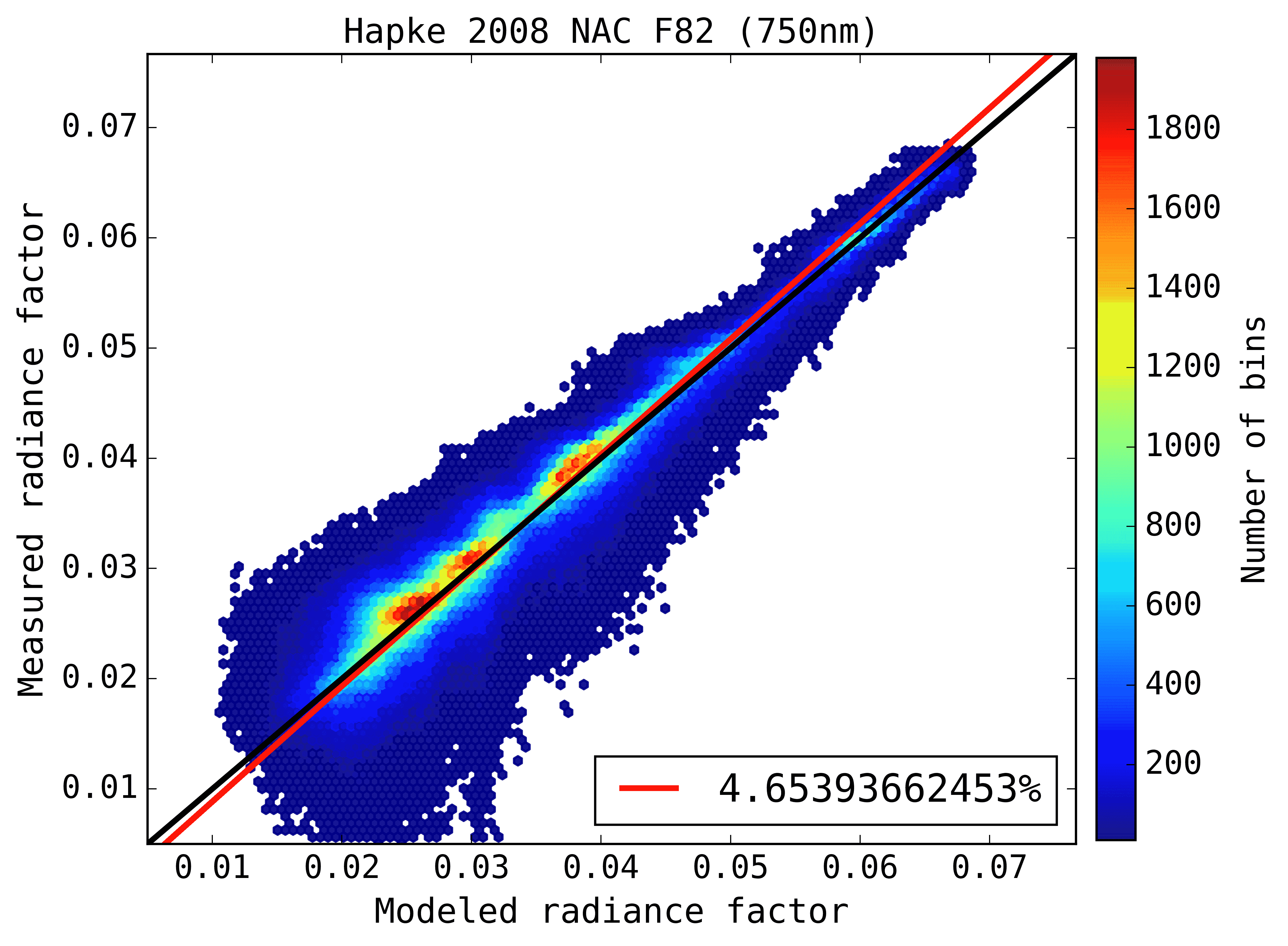}
	\subcaption{Goodness fit of the NAC F82 data}
	\includegraphics[scale=0.051]{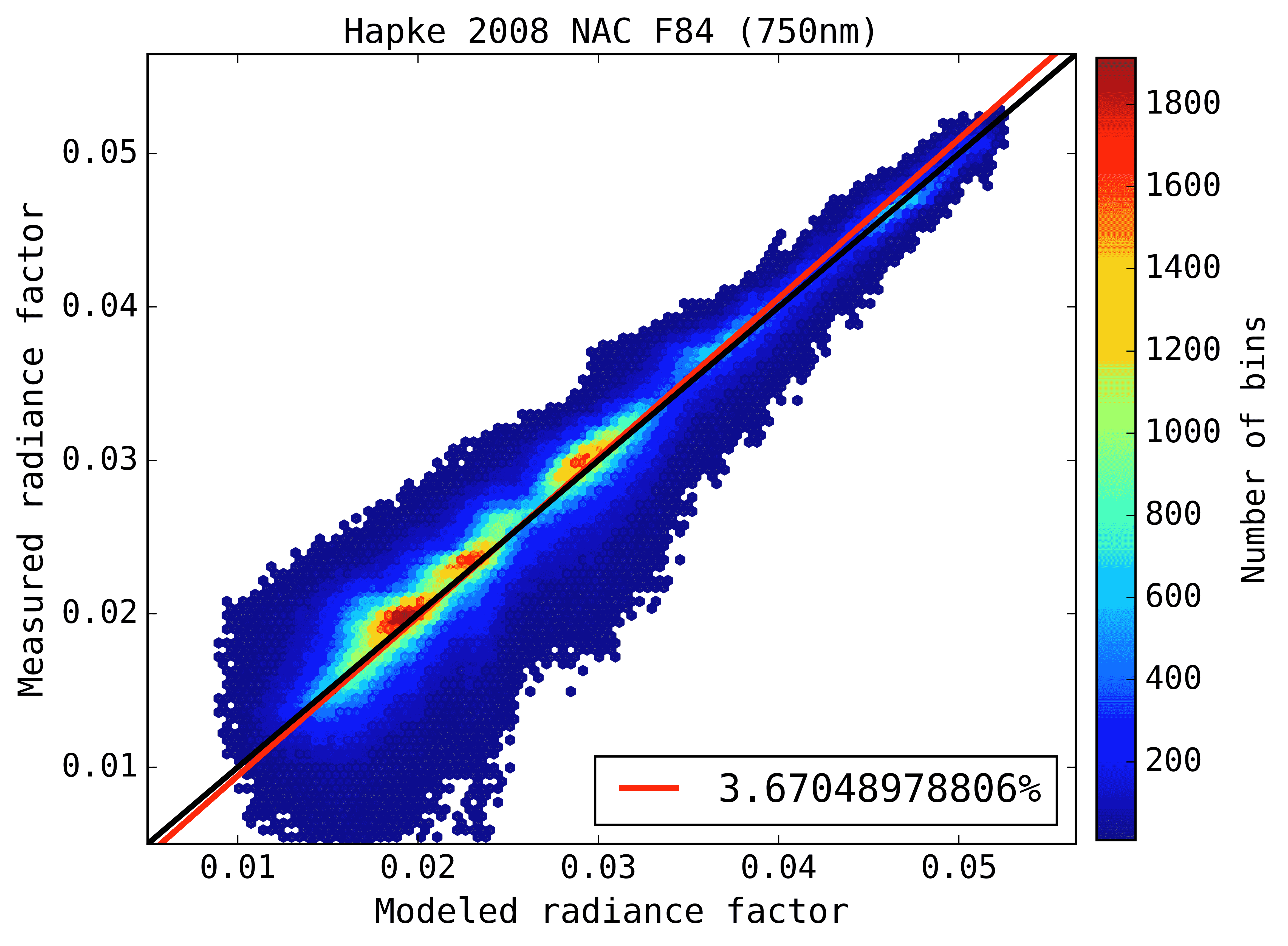}
	\subcaption{Goodness fit of the NAC F84 data}
	\includegraphics[scale=0.051]{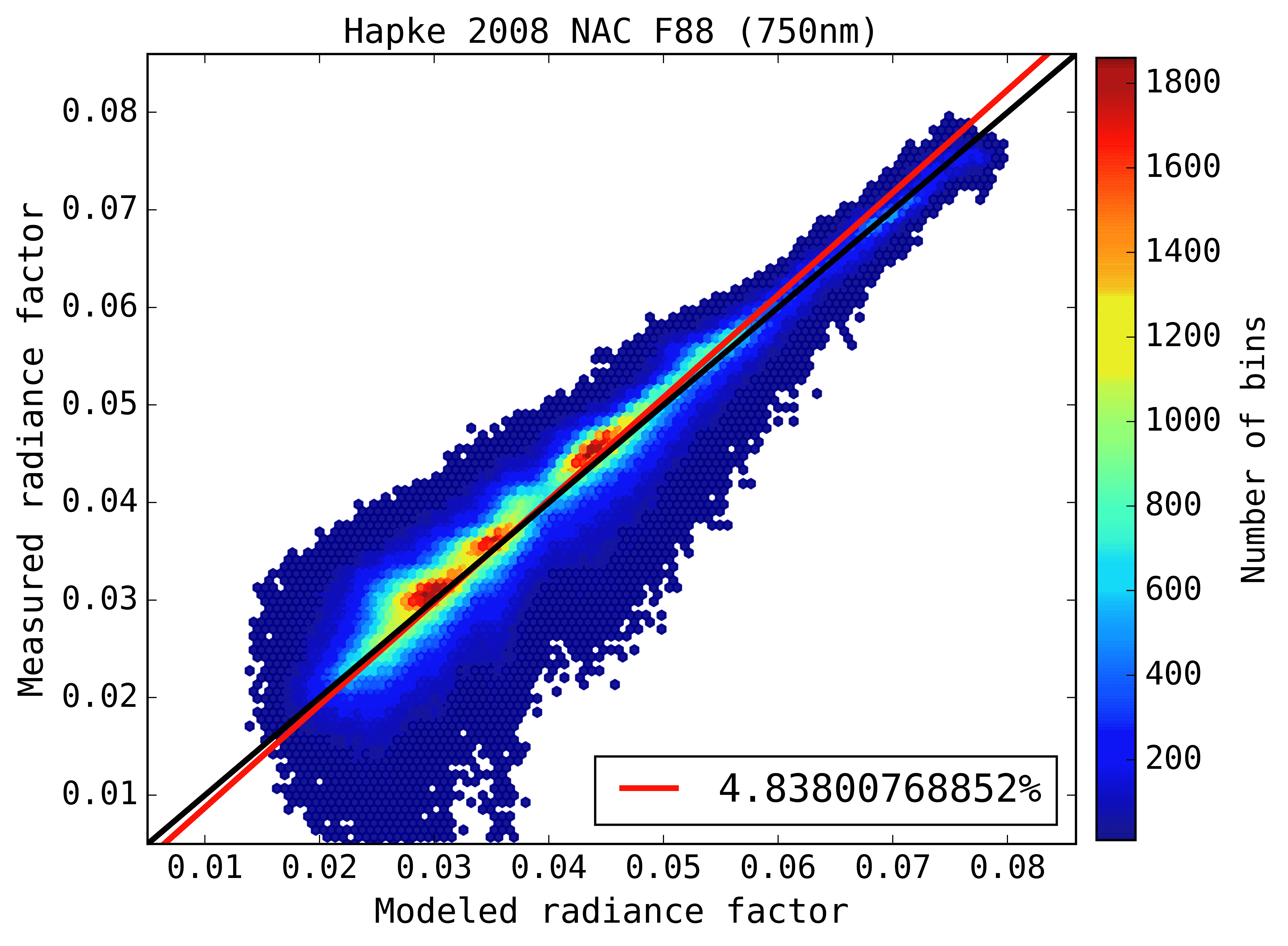}
	\subcaption{Goodness fit of the NAC F88 data}
\end{minipage}%
\begin{minipage}[b]{0.51\linewidth}
    \centering
	\includegraphics[scale=0.053]{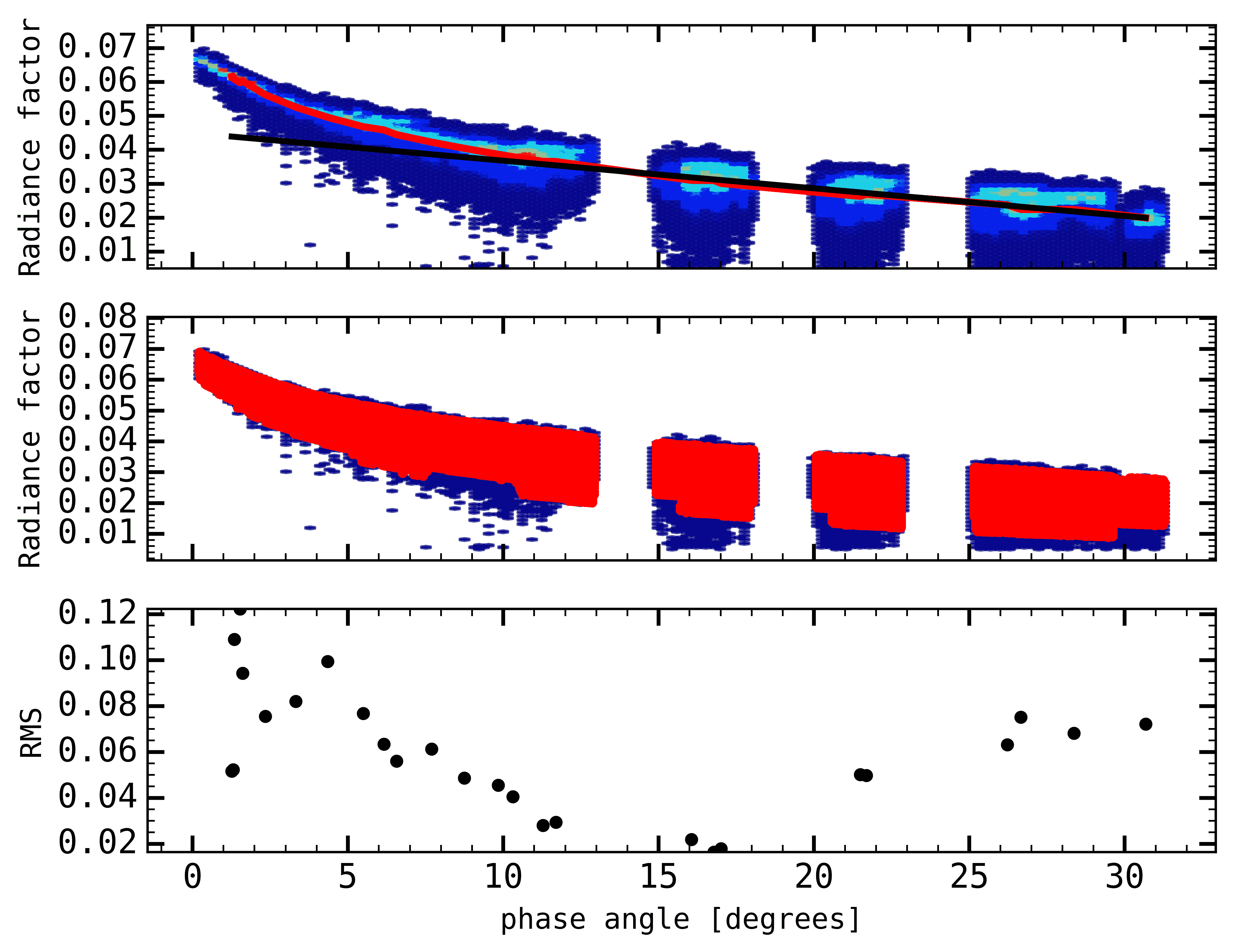}
	\subcaption{Fitting, simulation and corresponding RMS for the NAC F82 data}
	\includegraphics[scale=0.053]{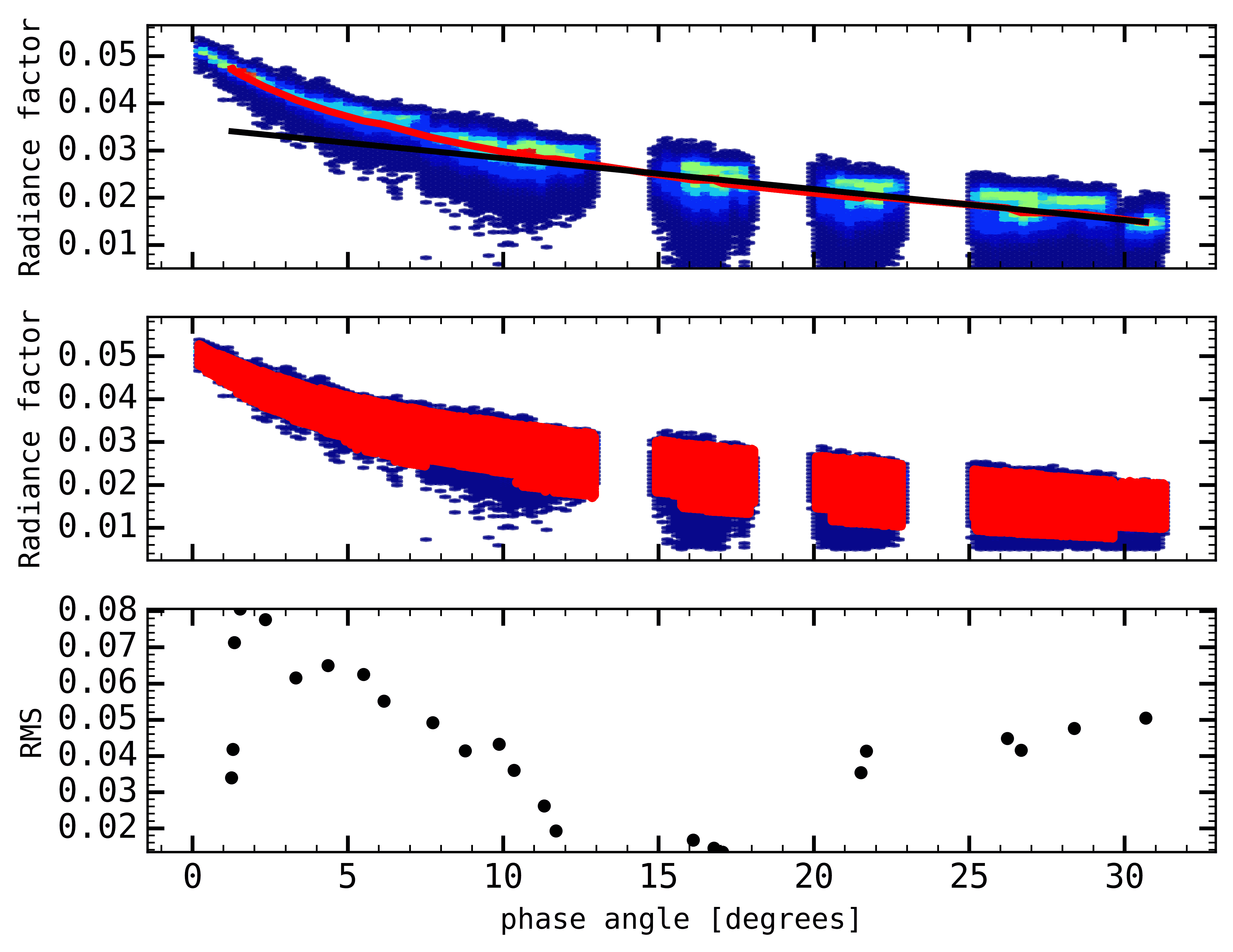}
	\subcaption{Fitting, simulation and corresponding RMS for the NAC F84 data}
	\includegraphics[scale=0.053]{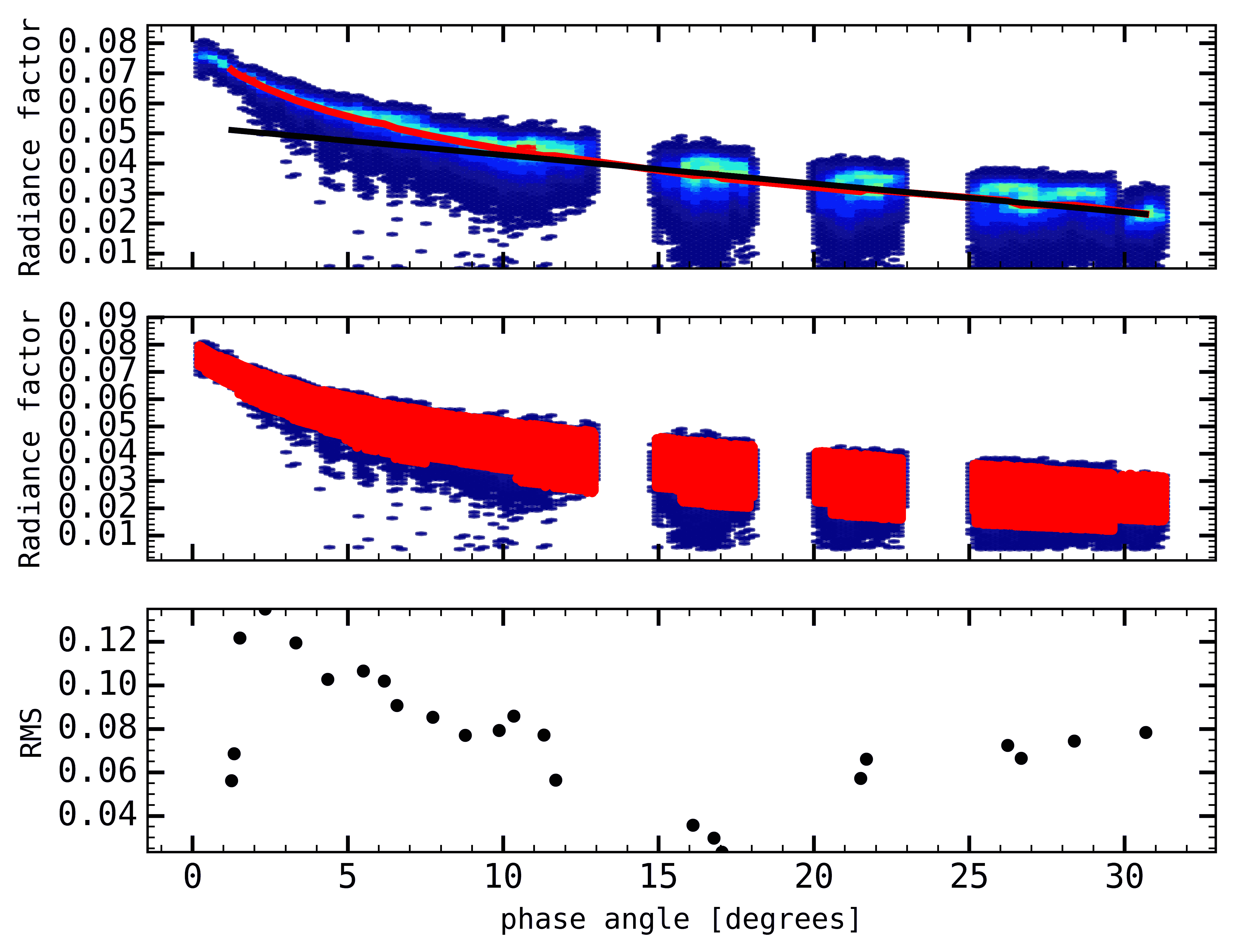}
	\subcaption{Fitting, simulation and corresponding RMS for the NAC F88 data}
\end{minipage}

\protect\caption{\label{fig:hapke-fit} Figures from the Hapke modelling. On the 
left-hand side, the 3 figures depicts for all 3 filters the measured RADF versus 
the modelled RADF. The black line represent a slope of 1 (i.e. a perfect match). 
The red line is our best linear fit by taking into account all active cells. 
On the right-hand side, the 3 panels of figures present the results of the 
photometric modelling for each NAC filter.
In each figure, the diagram at the top corresponds to the phase-function in the 
corresponding filter with the curve being the best-fit solution of the photometric 
modelling. The diagram in the middle presents an over-plot of the same phase-function 
with the simulation of the dataset using the 3D model and the best-fit solution. Finally, 
the diagram at the bottom display the mean of the RMS for each image. In all of these figures, as the measured radiance factor values have been binned, the first and second diagrams show a under-plotted density distribution
of the active cells.}
\end{figure*}

We fit to the three filters the five-parametric \citet{2002Icar..157..523H} model and 
HHS model. For all of them, the models have successfully converged to a single 
solution. Figure \ref{fig:hapke-fit} shows the observations together with the Hapke 
modelling for the three filters and the associated RMS. The total deviation of the 
modelled RADF compared to measured quantities (red line, Figures a-c-e) is no higher 
than $4.8\%$, reaching almost 1:1 ratio in respect to the small phase angles images 
(black line). The best Hapke parameters found for the different filters reproduce 
correctly variations due to illumination conditions.

The final Hapke solutions are shown in Table \ref{tab:Hapke-parameters}, together 
with pre-perihelion solutions obtained from the OSIRIS images from 21 July to 6 August 
2014 ($1.3^{\circ} < \alpha < 54^{\circ}$, \citealp{Fornasier_2015_AA_67PCG}). The parameters 
solution derived for the Ash-Imhotep area are very similar to those retrieved for the
whole northern hemisphere. The area has a very small single-scattering albedo, 
medium backscattering particle phase function, strong shadow-hiding mechanism, 
high top-layer porosity and smooth macroscopic roughness. \citet{2015A&A...583A..31C}, 
modelling the VIRTIS-M data of the 67P/CG comet for $27.2^{\circ}<\alpha<111.5^{\circ}$ 
from July 2014 to February 2015, obtained higher $w$ ({\small{}$0.06\pm0.01$}) and 
$g_{sca}$ (-0.42) values, but a $\overline{\theta}$ value consistent with our work. 
The discrepancies in the results between OSIRIS and VIRTIS data are related to the 
different phase angle range used to compute the phase curve slope, and to the different 
forms of Hapke modelling adopted.  Indeed, \citeauthor{2015A&A...583A..31C} neglected 
the opposition effect in their modelling.  This results in a steeper (more negative)  g$_{sca}$ 
value to account for the slope caused by opposition effect, and a steeper phase function 
that requires a higher single scattering albedo value to compensate for the overall 
measured I/F scale.

Rosetta provides for the first time the opportunity to derive the full set of parameters 
for a comet nucleus, particularly for the OE regime, as all the other cometary nuclei 
visited by spacecraft have been observed at $\alpha>11^{\circ}$. The surface analysed 
here and in \citet{Fornasier_2015_AA_67PCG} is characterised by a large $B_{s0}$
parameter, where according to \citet{9781139025683}, SHOE amplitudes higher than 
the nominal limit of 1.0 might indicate the presence of sub-particle structures, casting 
shadows onto one another, which would cause SHOE mechanism to be further enhanced. 
As most part of the top surface of 67P/CG is possibly covered by resettled fluffy dust layer 
in the low-gravity environment, it is not unpredictable that mutual shadow hiding 
among particles should not be an enhanced mechanism. Furthermore, the superficial 
porosity of the Ash-Imhotep region ($86\%$) is indistinguishable compared to that derived 
for the northern hemisphere. This high value is similar to the one determined by the Philae 
measurements for the first tens of metres during the first rebound on the comet
($75-85\%$, \citealp{Kofman_2015_Science_67P_Consert} ), and in agreement with the
porosity value of fractal aggregates \citep{Bertini_2007_AA_FractalAggregates, LevasseurRegourd_2007_PSS_CIPD, 
Lasue_2011_Icarus_LayeredNuclei} that are believed to be the best analogues of cometary dust. 
Low thermal inertia and high micro-roughness are proxy physical quantities to a high superficial 
porosity and a sharper SHOE. Laboratory experiments predict such correlation 
\citep{Zimbelman_1986_Icarus_Lavas,Piqueux_2009_JGRE_ThermCond_1}.
Therefore, as expected, our results are in agreement with high 
micro-roughness ($\overline{\theta_{r}}\approx55^{\circ}$, \citealp{Kamoun_2014_AA_67PArecibo}) 
obtained through re-analysis of radar echo from Arecibo observatory and the low 
therma inertia of $\sim85\ J\cdot m^{-2}\cdot K^{-1}\cdot s^{-1/2}$
provided by MUPUS instrument on-board Philae \citep{Spohn_2015_Science_MUPUS}.

Briefly, $w$, $g_{sca}$, $\overline{\theta}$ and $\rho_{v}$ are the only \citet{2002Icar..157..523H} 
parameters that can be compared to the literature. The Ash-Imhotep area along with all 
the northern hemisphere \citep{Fornasier_2015_AA_67PCG,2015A&A...583A..31C} show similar
photometric behavior to other nuclei in the visible spectral range, specially 9P/Tempel 1 
($w_{650nm}=0.039$, $g_{sca}=-0.49$, $\overline{\theta}=16^{\circ}\pm8^{\circ}$
and $\rho_{v}=0.059$, \citealp{2007Icar..187...41L}) and 103P/Hartley 2 
($w_{750nm}=0.036$, $g_{sca}=-0.46$, $\overline{\theta}=15^{\circ}\pm10^{\circ}$
and $\rho_{v}=0.045$,  \citealp{2013Icar..222..559L}). The geometric albedo $\rho_{v}$ of the 
67P/CG nucleus has an intermediate value compared to that found for 81P/Wild 2 
($\rho_{v}=0.059$,\citealp{2009Icar..204..209L}) and 19P/Borrelly ($\rho_{v}=0.072\pm0.002$, 
\citealp{Li_2007_Icarus_19PB_Photometry}). Moreover, 67P/CG shows $g_{sca}$ lower than all other nuclei. None of
previous spacecraft comets attained the same spatial resolution than Rosetta in their cometary 
images. However, both objects that present similar Hapke parameters, Tempel 1 and Hartley 2, are 
those that also demonstrate overflow of extremely smooth and fine material in their surface, in a 
comparable manner to the Hapi region on 67P/CG.

\begin{center}
\begin{table*}
\begin{centering}
\begin{tabular}{lc>{\centering}p{1.8cm}>{\centering}p{1.5cm}>{\centering}p{1.5cm}>{\centering}p{1.5cm}>{\centering}p{1.2cm}>{\centering}p{1.5cm}c>{\centering}p{1cm}}
\hline 
{\small{}Model} & {\small{}Filter} & \emph{\small{}$w_{\lambda}$} & 
\emph{\small{}$g_{sca_{\lambda}}$} & \emph{\small{}$B_{SH,0}$} & 
\emph{\small{}$h_{SH}$} & \emph{\small{}$\overline{\theta}[deg]$} & 
\emph{\small{}$K$} & {\small{}Porosity} & {\small{}$p_{v}$}\tabularnewline
  &  &  $(\pm0.001)$ &  $(\pm0.02)$ & $(\pm0.3)$ & $(\pm0.005)$ & $(\pm0.2)$ & $(\pm0.005)$ 
&  & \tabularnewline  \hline
\hline
{\small{}H2002$^{a}$} & {\small{}F82} & {\small{}$0.042$} & {\small{}$-0.37$} & {\small{}$2.5$} & {\small{}$0.079$} & {\small{}$15$} & {\small{}-} & {\small{}-} & {\small{}$0.064$}\tabularnewline
\hline 
\hline
{\small{}H2002$^{\dagger}$} & {\small{}F84} & {\small{}$0.032$} & {\small{}$-0.39$} & {\small{}$2.57$} & {\small{}$0.067$} & {\small{}$16.5$} & {\small{}-} & {\small{}-} & {\small{}$0.052$}\tabularnewline
\hline 
{\small{}H2002} & {\small{}F82} & {\small{}$0.046$} & {\small{}$-0.37$} & {\small{}$2.56$} & {\small{}$0.064$} & {\small{}$15.6$} & {\small{}-} & {\small{}-} & {\small{}$0.068$}\tabularnewline
\hline 
{\small{}H2002} & {\small{}F88} & {\small{}$0.053$} & {\small{}$-0.36$} & {\small{}$2.52$} & {\small{}$0.064$} & {\small{}$15.1$} & {\small{}-} & {\small{}-} & {\small{}$0.079$}\tabularnewline
\hline 
\hline
{\small{}HHS$^{a}$} & {\small{}F82} & {\small{}$0.034$} & {\small{}$-0.42$} & {\small{}$2.25$} & {\small{}$0.061$} & {\small{}$28$} & {\small{}$1.20$} & {\small{}$0.87$} & {\small{}$0.067$}\tabularnewline
\hline 
\hline
{\small{}HHS$^{\ddagger}$} & {\small{}F84} & {\small{}$0.026$} & {\small{}$-0.39$} & {\small{}$2.56$} & {\small{}$0.067$} & {\small{}$16.3$} & {\small{}$1.226$} & {\small{}$0.85$} & {\small{}$0.052$}\tabularnewline
\hline 
{\small{}HHS} & {\small{}F82} & {\small{}$0.038$} & {\small{}$-0.37$} & {\small{}$2.57$} & {\small{}$0.064$} & {\small{}$15.6$} & {\small{}$1.219$} & {\small{}$0.86$} & {\small{}$0.068$}\tabularnewline
\hline 
{\small{}HHS} & {\small{}F88} & {\small{}$0.044$} & {\small{}$-0.36$} & {\small{}$2.64$} & {\small{}$0.066$} & {\small{}$15.2$} & {\small{}$1.212$} & {\small{}$0.85$} & {\small{}$0.079$}\tabularnewline
\hline

 &  &  &  &  &  &  &  &  & \tabularnewline
\end{tabular}
\par\end{centering}

\begin{raggedright}
{\scriptsize{}$(\dagger)$ Hapke 2002 model.}
\par\end{raggedright}{\scriptsize \par}
\begin{raggedright}
{\scriptsize{}$(\ddagger)$ Helfenstein-Hapke-Shkuratov model (see section \ref{Subsection:  HHS reference} ).}
\par\end{raggedright}{\scriptsize \par}
\begin{raggedright}
{\scriptsize{}$(a)$ Obtained by \citep{Fornasier_2015_AA_67PCG}, from OSIRIS data}
\par\end{raggedright}{\scriptsize \par}
\caption{\label{tab:Hapke-parameters}Hapke parameters for the February 2015 flyby area.}
\end{table*}
\par\end{center}

\subsection{Investigation of bright patches and dark boulders}
\begin{figure}
\centering
\includegraphics[scale=0.22]{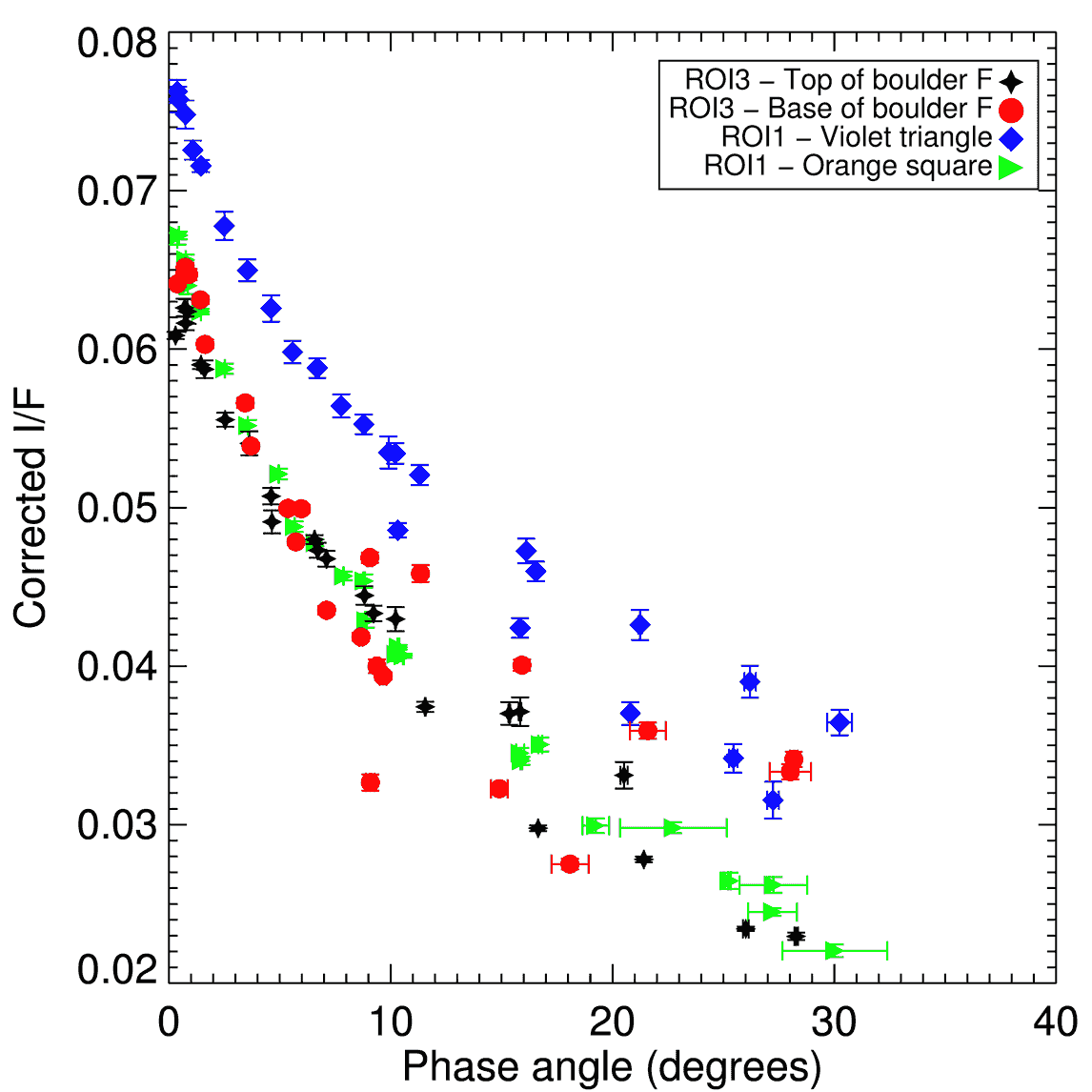}
\caption{Phase curves at 649nm of a bright spot (the violet triangle in ROI1), of a dark boulder (green star in ROI3) and their respective contiguous terrains (orange square in ROI1) and the base of feature D in ROI3 (its position is indicated with the blue ellipse in Fig. \ref{panel: ID65-12h19}).}
\label{fig: Phase functions}
\end{figure}

As mentioned in section \ref{Section 4}, the Fig. \ref{fig: Phase functions} shows the phase curves of four selected areas which have different morphology, including a sombre boulder and a bright spot.
The bright spot investigated here corresponds to the violet triangle in ROI1, while the sombre boulder corresponds to the 
feature D in ROI3. As expected, we observe that the reflectance of this feature, which 
appears bright under every illumination conditions in every image, is larger  
than that of the dark boulder. We also note, for the bright spot, that 
the opposition effect surge at small phase angle is not pronounced as what could be 
expected in the case of water-ice rich material. We also report that the dimensions 
of the dark boulder are only merely above the vertical and horizontal accuracy of the 
3D model of the region: hence though this feature is visible in the 3D model, its 
reconstitution lead to a disparity in the generated phase angles.

\begin{table*}
\centering
\begin{tabular}{|c|c|c|c|c|c|c|}
\hline
Feature & w$_{649nm}$ & g$_ {SCA}$ & B$_{SH,0}$ & h$_{SH}$ & Porosity & $p_{v}$ \\
\hline
\hline
Bright spot (ROI 1)& 0.068$\pm$0.002 & -0.31$\pm$0.02 & 1.583$\pm$0.25 & 0.089$\pm$0.01 & 0.81$\pm$0.02 & 0.077\\
\hline
Sombre boulder (ROI 3)& 0.0385$\pm$0.002 & -0.45$\pm$0.03 & 1.526$\pm$0.41 & 0.087$\pm$0.01 & 0.82$\pm$0.02 & 0.064\\
\hline
\hline
\end{tabular}
\caption{Results of HHS modelling for the surface elements investigated}
\label{Table:HHS - Fast Hapke model}
\end{table*}

The Hapke parameters for the bright spot and the dark boulder present a clear 
dichotomy between the two objects (see Table \ref{Table:HHS - Fast Hapke model}). The bright spot is characterized by a high 
normal and single-scattering albedo and a flatter phase curve slope (g$_{sca}$). 
The reverse is measured for the dark boulder. As the phase curve slope is inversely
proportional to the geometric albedo \citep{Deau_2013_Icarus_OE_labexps}, the 
same SHOE width and amplitude might indicate that the micro-roughness structure 
in the top layers is preserved.

\begin{figure}
  \centering
   \includegraphics[scale=0.109]{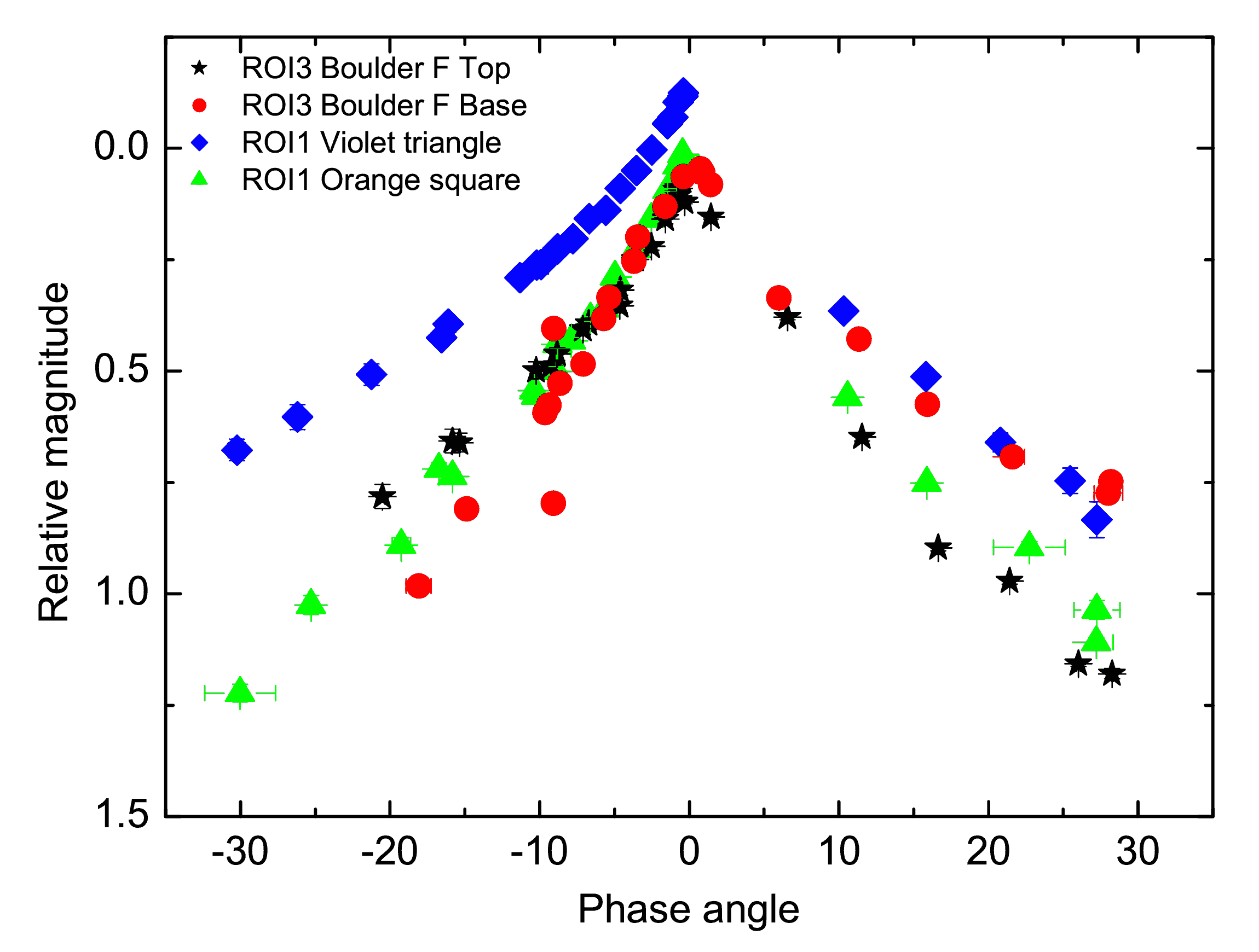}
   \caption{Diagram of the difference in magnitude for the phase curves of the 4 selected areas (cf. Fig. \ref{fig: Phase functions}), when normalised to the averaged comet albedo (6.8 \%).}
   \label{fig:Irina1}
\end{figure}

 The radiance factor values normalized to the averaged comet albedo (0.068) and converted to magnitudes, are presented in Fig. \ref{fig:Irina1}.
 One of the considered areas (violet triangle at ROI1) is noticeably brighter while other three areas have similar albedos. 
 An asymmetry of phase curves measured before and after opposition is well seen. It becomes evident at a phase angle larger than 10-15$^{\circ}$ and can be most probably explained by variations of local elevation. 
 The Ash terrain, which is a more plane surface, shows the smallest difference between phase curves before and after opposition.

To compare phase behaviours of selected areas, we considered only phase angles up to 10 deg where an influence of the relief is less significant. 
The brighter region exhibits a distinct opposition effect as contrary to darker regions, for which the magnitude-phase dependences are linear at small phase angles.
The linear phase slopes are 0.045$\pm$0.002 mag/deg (green star at the top of boulder D), 0.062$\pm$0.005 mag/deg (blue ellipse at the feet of boulder D) and 0.051$\pm$0.002 mag/deg (orange square at the base of the bright spot in ROI1). 
The differences in phase slopes are most probably related to the variation of surface roughness. 
The boulder which looks darker than its surroundings should have considerably smoother surface compared to the surface at its base.\\

The values of phase slopes found for selected areas of the comet are within typical 
values observed for comets (e.g. \cite{Li_2007_Icarus_19PB_Photometry}), and in particular similar to 
what has been found for the global 67P/CG (0.047$\pm$0.002 mag/deg for 
$\alpha >$7$^{\circ}$) from early resolved images of the nucleus \citep{Fornasier_2015_AA_67PCG}, 
and in particular for the sombre boulder and the smooth Ash terrain.

\subsection{Comparison to radio-goniometric measurements}
The February 2015 flyby provides the first-time opportunity to retrieve and analyse  
photometric parameters at spatial resolution comparable to those obtained through 
laboratory experiments. Recently, in the framework of Rosetta mission, new spectro-photometric 
measurements on cometary analogues were performed at the University of Bern using 
the PHIRE-2 radio-goniometer and the SCITEAS simulation chamber 
\citep{Jost_2016_Icarus_LabExp_Opp, 2011P&SS...59.1601P, 2015P&SS..109..106P, Pommerol_2015_AA_ExposedIce}.
The PHIRE-2 radio-goniometer measured the scattering curve before 
and  after sublimation of inter/intra-mixture samples of $67\pm31\mu$m
ice particles, 0.2 wt\% carbon black and 0.1 wt\% tholins at 750 nm. The inter- and 
intra- mixtures are produced using different sample preparation protocols and 
differ by the structure of the sample at the scale of the individual grains.  
The dust and water-ice grains are individually separated in the inter-mixture, 
whereas minerals grains are contained inside the water-ice grains for intra-mixture 
preparations. Each experiment was set up into a $2\times4\times2$ cm$^{3}$ sample holder, 
providing 4 combinations of cometary analogue surfaces plus an ice-less combination of  
tholins (33\%) and carbon black (66\%), we refer the reader to \citet{Poch_2016_Icarus_IceTholins} 
for more details on the preparation methods.
The resolution of comet data from the 2015-February fly-by is of 11cm/pxl, whereas
the analogue surfaces are about only 10 times smaller, placing them as
an equivalent to the highest-resolved surface unit of 67P/CG up to now. 

\begin{figure*}
\begin{minipage}[b]{0.49\linewidth}
    \centering
	\includegraphics[scale=0.068]{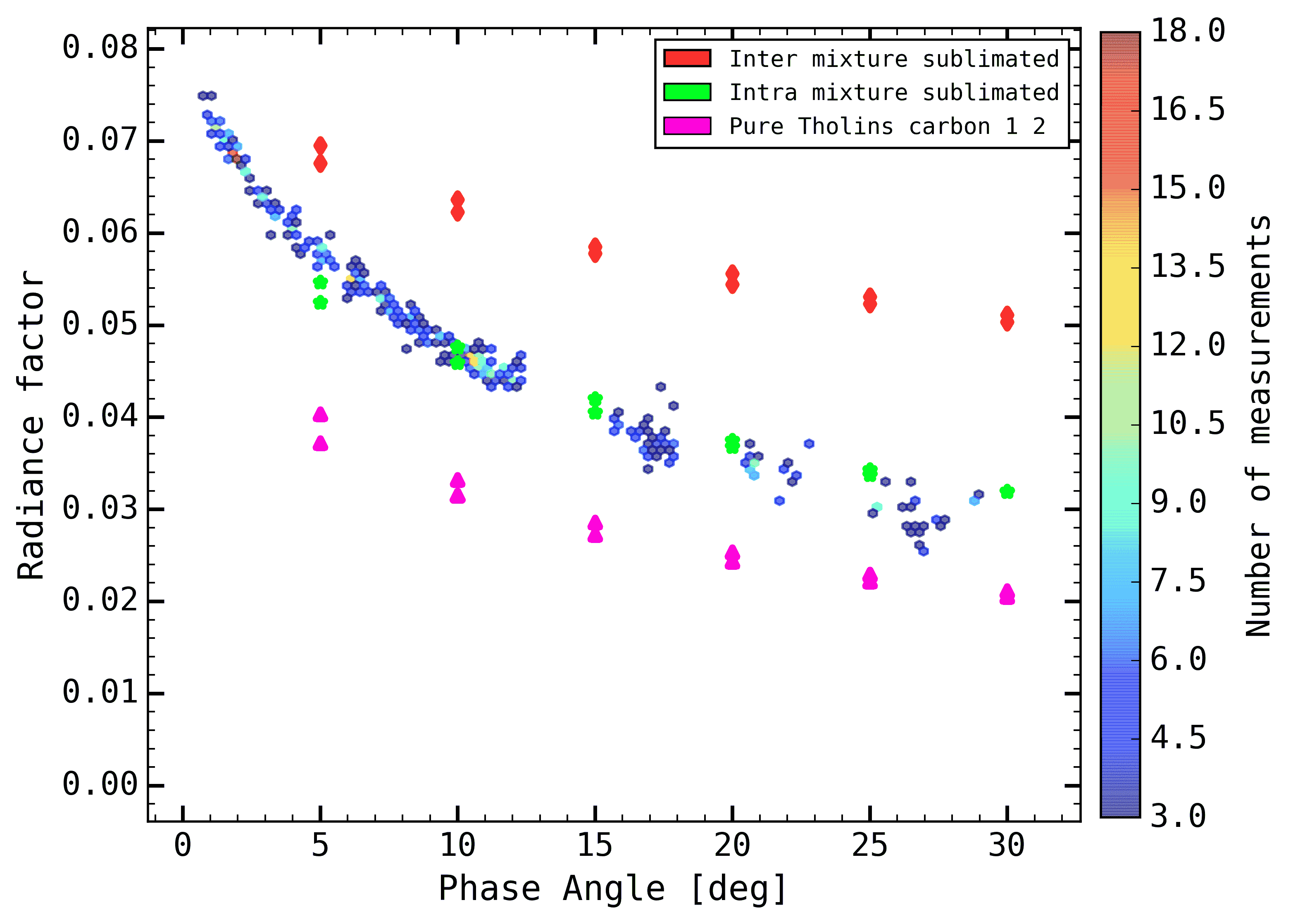}
	\subcaption{laboratory experiments comparison}
\end{minipage}%
\begin{minipage}[b]{0.49\linewidth}
    \centering
	\includegraphics[scale=0.045]{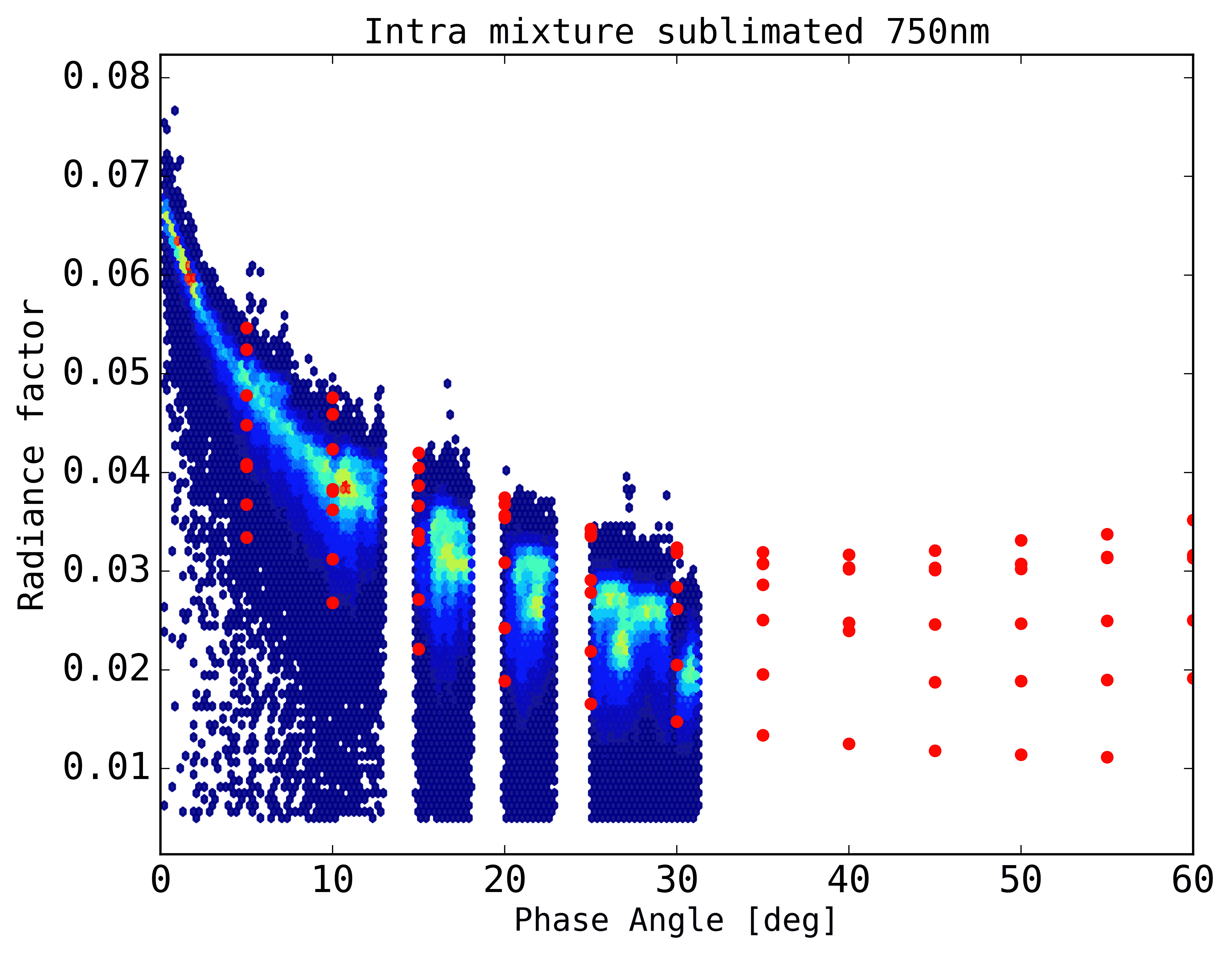}
	\subcaption{laboratory experiments comparison}
\end{minipage}
\caption{\label{fig:lab-comparison} (a) Comparison phase curves of dry laboratory
samples and NAC F88 Fly-by area for phase angle smaller than 35 degrees
at 750 nm. For both data, it was selected measurements at incidence
angle equal to zero. The dry samples are the sublimated intra- and
inter-mixtures, and an assemblage of tholins (33\%)+carbon black(66\%). (b)
Best Match: Sublimated intra-mixture. The scatter of measurements
well represent the disk behaviour observed in the fly-by phase curve.
The lack points at $\alpha<5^{\circ}$do not allow to infer any information
about the opposition surge.}
\end{figure*}

As the cometary surfaces investigated here are essentially ice-free, we compare 
the photometric properties derived from NAC F88 flyby data to laboratory data 
obtained for the fully sublimated state of 3 different types of cometary surfaces
analogues (Jost et al., in preparation). The laboratory reflectance data were measured 
with the PHIRE-2 radio-goniometer at a central wavelength of 750 nm. The comparison, 
as presented in Figure \ref{fig:lab-comparison}a for incidence angle equal to zero, 
clearly indicates that the sublimated “intra-mixture” sample
(see Fig. \ref{fig:lab-comparison}b) displays a similar scattering curve in the phase 
angle range of the fly-by. The three samples show quite similar phase curve slope, 
whereas inter-mixture is slightly flatter than intra-mixture for $\alpha>15^{\circ}$.
Nonetheless, the intra-mixture is the only sample to fit the 67P/CG phase curve in absolute radiance factor and
also in slope.

At larger phase angles ($\alpha\gtrsim40^{\circ}$), the intra-mixture scattering 
curve becomes wider and strongly forward scattering, which is very different 
from what is observed by \cite{Ciarniello_2015_AA_67P_VirtisM} for $\alpha>80^{\circ}$, 
where the overall reflectance goes fainter. This is essentially due to the effect 
of large scale topography and roughness that are absent from the laboratory 
samples but dominate the photometry of the nucleus at these large phase angle. 
Moreover, radiometric measurements were not obtained to $\alpha<5^{\circ}$,
which does not allow us to infer the shape and magnitude of the opposition
surge.

According to \citet{Poch_2016_Icarus_Mixture_sublimations}, an intra-mixed surface has
as characteristic the formation of mantles of high internal cohesiveness
and with episodic and bigger release of fragments during sublimation
than inter-mixed counterpart. Under the optical coherence tomographic
scan (Jost et al., in preparation), intra-mixtures also presents a higher superficial
roughness than inter-mixtures, which would also indicate an higher
superficial porosity, as found by the photometric modelling of the
comet. However, none of these surface properties have been quantitatively
characterized yet.

\section{Discussions and outlook}

The analysis of the decimetre-scaled observations of the comet 67P/CG 
with the OSIRIS instrument reveals that the photometric parameters of the flown-by 
region are similar to those of the whole nucleus. The comet surface is noticeably dark, 
its uppermost layer is very porous, and it preferentially backscatters the reflected light 
(see Table \ref{tab:Hapke-parameters}).
The spectrophotometric analysis of smooth and dusty area on both Ash and Imhotep regions  
indicate similar properties in terms of texture and composition. These terrains are found to 
be among the reddest in term of spectral slope on the comet surface. 
We note here that the texture of the material covering the Ash part of the flown-by region
appear similar to the observations of the ROLIS instrument on-board the Philae lander, 
as presented in \cite{Mottola_2015_Science_RolisDescent}. We also observe similar textures in
parts of the flown-by region: a smooth regolith with the presence of decimetre and 
metre-sized boulders, alone or as part of clusters, and alcoves.

However, the spectrophotometric analysis also points out to local variations in terms of
reflectance and colours (respectively up to 40\% and 50\%) pointing to local heterogeneities 
at the metre and decimetre scale associated to peculiar morphological features such as boulders.
Heterogeneities were confirmed at larger scales by OSIRIS and at smaller scales 
by CIVA onboard Philae \citep{Bibring_2015_Science_CIVA}, both in terms of texture and composition.

In their study of the Agilkia region (where the Philae lander made its first touch 
down), \cite{LaForgia_2015_AA_Agilkia} found that the area closest 
to outcrops and cliffs were those exhibiting the reddest spectral behaviour, which is in 
agreement with our observations of Fig. \ref{panel: spectral slopes}.
Furthermore, they also investigated the spectrophotometric difference between fine 
deposits and outcrops area, and they found that fine deposits have a relatively bluer 
spectrum than outcrops, a result which is in agreement with our observations in the different ROIs.

The presence and origin of fine deposits in the Agilkia and Ash regions have already 
been investigated and discussed in \citet{Thomas_2015_Science_67P_morphology, LaForgia_2015_AA_Agilkia}:
the authors speculate that those fine deposits might be the results of airfall of material 
originating from other active regions, such as Hapi.

The investigation of the origin of fine deposits in the flyby-region is beyond 
the scope of this paper, yet we can retain that an airfall hypothesis could explain why 
the top of outcrops and boulders have a spectral behaviour similar to pebble-less neighbouring surfaces.

When investigating  the median spectral slope of the area flown-by, and the 
spectra of large boulder-less surfaces (1.5x1.5m$^{2}$), as in \cite{LaForgia_2015_AA_Agilkia}, 
we found likewise that the surface was revealing a global red spectral behaviour 
(S$_{med}$ = 17.7\%/100nm at 1$^{\circ}$ of phase angle) without any evidence of 
band features (with the exception of the cometary emission over-flux) in the near-UV to near-IR domain.
According to VIRTIS results, the red spectral behaviour is compatible with an organic-rich composition 
\citep{Capaccioni_2015_Science_Virtis, Filacchione_2016_Icarus_GlobalSurfaceVirtis}. 
 
\begin{figure*}
\centering
\includegraphics[scale=0.25]{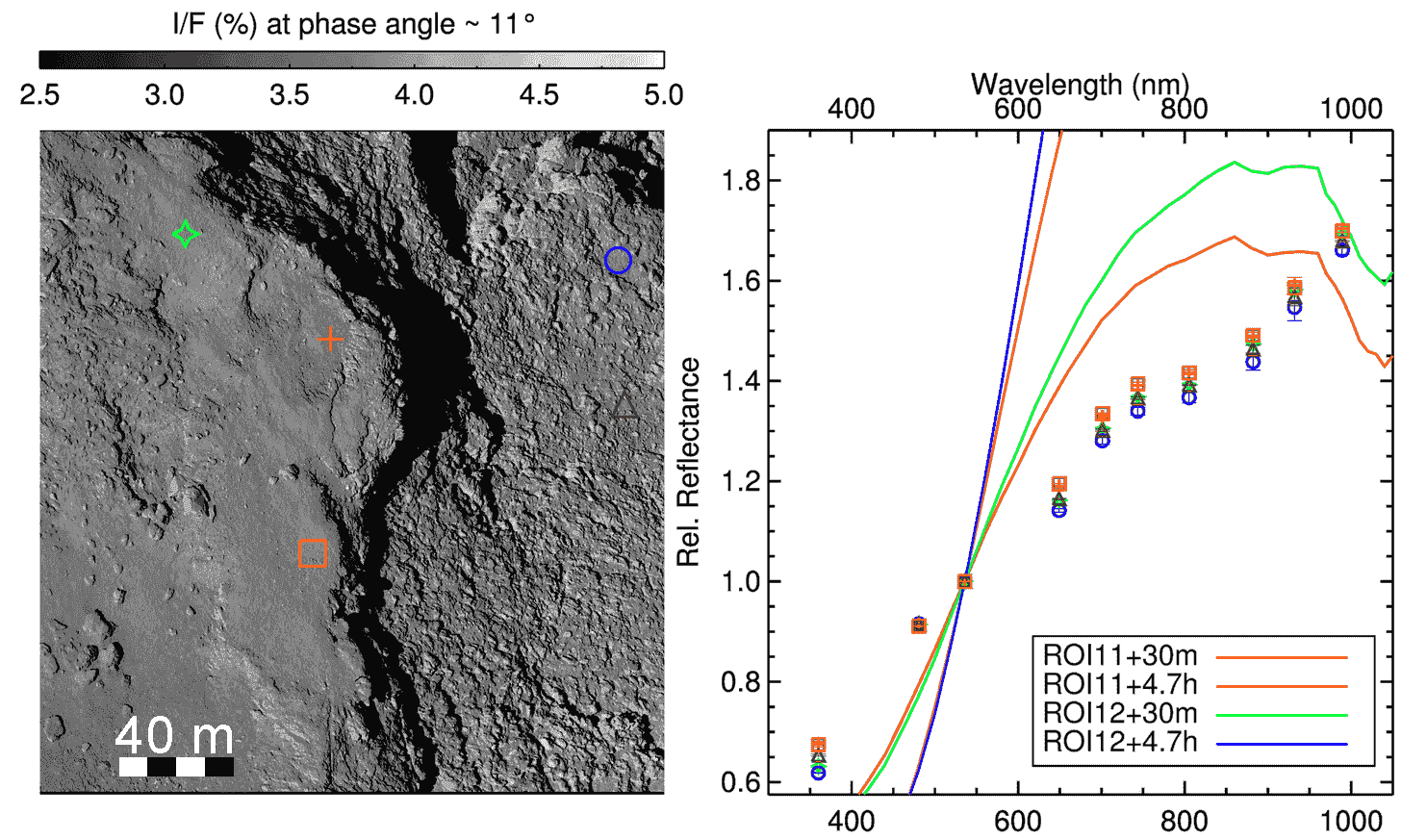}
\caption{Comparison with laboratory measurements: the ROIs of the legend refer to those defined in \citet{Poch_2016_Icarus_Mixture_sublimations}, while the time indicated refers to the lapse of time the samples were left to sublimate.}
\label{fig: comp bern}
\end{figure*}

When comparing our dataset to \citet{Jost_2016_Icarus_LabExp_Opp}, we found that a 
intra-mixture composed of tholins and of carbon black was a best-match in terms 
of photometric properties. However, when comparing spectra of fresh and sublimated 
samples of intra-mixture and inter-mixture of water-ice, tholins and minerals (olivine and smectite), 
we found that no sample would match well the spectrophotometric properties of 
the surface layer of the flown-by region, as depicted in Fig. \ref{fig: comp bern}.\\
Reflecting that such samples are not a proxy for the comet's composition, we can 
nevertheless remark that the after-sublimation intra-mixture sample was a better 
match than the inter-mixture sample in terms of photometric properties.

As we have previously mentioned, in this region, the bright spots display a steep 
spectral slope value, which is not normally associated to terrains enriched in water-ice on the comet 67P/CG.
Moreover, the reflectances of the bright spot is not as high (3 to 8 times the reflectance of the 
surroundings) as normally observed in exposed water-ice spots on the comet
\citep{Pommerol_2015_AA_ExposedIce, Barucci_2016_AA_Ices_Features,Fornasier_2016_Science_IceDust}.

All of the observed boulders of the flown-by region showing a lower reflectance at small phase angle, 
relative to their surroundings, also exhibit a red spectral behaviour, often much redder than the 
surroundings. However, also brighter regions in or close to the dark boulders, like the one observed 
on boulders A or F often show a red spectral behaviour very close to that of the 
darker boulders. This indicates that the uppermost composition must be similar 
in the aforementioned structures, and that the different reflectance may be related 
to grain size properties, with brightest features having probably smaller grain size.\\

Yet, those measurements give an extraordinary insight at the decimetre-scaled 
structure of the nucleus and surely a preview of the upcoming OSIRIS observations 
as the ROSETTA spacecraft will orbit, in the coming months, closer and closer to the nucleus 
of comet 67P/Churyumov-Gerasimenko.

\section*{Acknowledgements}
OSIRIS was built by a consortium of  the Max-Planck-Institut f\"ur Sonnensystemforschung, G\"ottingen, Germany, CISAS--University of Padova, Italy, the Laboratoire d'Astrophysique de Marseille, France, the Instituto de Astrof\'isica de Andalucia, CSIC, Granada, Spain, the Research and Scientific Support Department of the European Space Agency, Noordwijk, The Netherlands, the Instituto Nacional de T\'ecnica Aeroespacial, Madrid, Spain, the Universidad Polit\'echnica de Madrid, Spain, the Department of Physics and Astronomy of Uppsala University, Sweden, and the Institut  f\"ur Datentechnik und Kommunikationsnetze der Technischen Universit\"at  Braunschweig, Germany. The support of the national funding agencies of Germany (DLR), France (CNES), Italy (ASI), Spain (MEC), Sweden (SNSB), and the ESA Technical Directorate is gratefully acknowledged.\\
Rosetta is an ESA mission with contributions from its member states and NASA. Rosetta's Philae lander is provided by a consortium led by DLR, MPS, CNES and ASI.\\
The SPICE, SPICE/DSK libraries and PDS resources are developed and maintained by NASA. The authors wish to thank Nathan Bachmann and his colleagues for their guidance.\\




\bibliographystyle{mnras}
 \bibliography{20150214_Spectro_Photo_V10}



\appendix




\bsp	
\label{lastpage}
\end{document}